\documentclass[11pt]{article}
\usepackage{amsmath,amssymb,graphicx}
\usepackage{breakurl} 
\usepackage{hyperref}
\usepackage{multicol,color,longtable}
\definecolor{darkred}{rgb}{0.65,0.15,0}
\hypersetup{pdfborder={0 0 0},colorlinks=true,urlcolor=blue,citecolor=blue,linkcolor=darkred,linktocpage=true}

\usepackage{cite}
\usepackage{array,multirow}

\newcommand{\eprint}[1]{{\href{http://arxiv.org/abs/#1}{[\texttt{#1}]}}}
\newcommand{\eprintN}[1]{{\href{http://arxiv.org/abs/#1}{[\texttt{#1 [hep-th]}]}}}
\newcommand{\eprintNT}[1]{{\href{http://arxiv.org/abs/#1}{[\texttt{#1 [math-NT]}]}}}
\newcommand{\doi}[2]{{\hypersetup{urlcolor=darkred}\href{http://dx.doi.org/#2}{#1}\hypersetup{urlcolor=blue}}}

\newcommand{\nn}{\nonumber}

\newcommand{\gs}{g_{\mathrm{s}}}
\renewcommand{\Im}{\mathrm{Im}}
\renewcommand{\Re}{\mathrm{Re}}

\newcommand{\ints}{\mathbb{Z}}
\newcommand{\reals}{\mathbb{R}}
\newcommand{\UHP}{\mathbb{H}}
\newcommand{\Li}{\textrm{Li}}

\makeatletter

\@addtoreset{equation}{section} \makeatother

\begin{document}
\setcounter{page}{0}

{\flushright {DCPT-19/01}\\[25mm]}

\begin{center}
{\LARGE \bf Modular graph functions\\[4mm] and asymptotic expansions of Poincar\'e series}\\[10mm]

\vspace{8mm}
\normalsize
{\large  Daniele Dorigoni${}^{1}$ and Axel Kleinschmidt${}^{2,3}$}

\vspace{10mm}
${}^1${\it Centre for Particle Theory \& Department of Mathematical Sciences\\
 Durham University, Lower Mountjoy, Stockton Road, Durham DH1 3LE, UK}
\vskip 1 em
${}^2${\it Max-Planck-Institut f\"{u}r Gravitationsphysik (Albert-Einstein-Institut)\\
Am M\"{u}hlenberg 1, DE-14476 Potsdam, Germany}
\vskip 1 em
${}^3${\it International Solvay Institutes\\
ULB-Campus Plaine CP231, BE-1050 Brussels, Belgium}

\vspace{20mm}

\hrule

\vspace{10mm}

\begin{tabular}{p{13cm}}
{\small
In this note we study $SL(2,\mathbb{Z})$-invariant functions such as modular graph functions or coefficient functions of higher derivative corrections in type IIB string theory. The functions solve inhomogeneous Laplace equations and we choose to represent them as Poincar\'e series. In this way we can combine different methods for asymptotic expansions and obtain the perturbative and non-perturbative contributions to their zero Fourier modes. In the case of the higher derivative corrections, these terms have an interpretation in terms of perturbative string loop effects and pairs of instantons/anti-instantons.
}
\end{tabular}
\vspace{7mm}
\hrule
\end{center}

\thispagestyle{empty}

\newpage

\setcounter{page}{1}
\setcounter{tocdepth}{2}
\tableofcontents

\vspace{5mm}
\hrule
\vspace{5mm}

\section{Introduction}

\subsection{Two instances of \texorpdfstring{$SL(2,\ints)$}{SL(2,Z)} in string theory}

$SL(2,\ints)$ automorphic forms and functions arise in closed string theory in at least two distinct instances, depending on the interpretation of the modular group $SL(2,\ints)$. In the first instance, $SL(2,\ints)$ plays the role of the mapping class group of the toroidal genus-one world-sheet and is thus associated to perturbative aspects of the string at one-loop order and the group $SL(2,\ints)$ acts on the modular parameter $\tau$ of the string world-sheet. The second instance is when $SL(2,\ints)$ is the non-perturbative U-duality group of the type IIB string in ten dimensions, and we shall now describe both cases in more detail.

The appearance of $SL(2,\ints)$ automorphic forms in closed string scattering at one-loop order has been recently formalised in the framework of \textit{modular graph functions}~\cite{DHoker:2015gmr} and \textit{modular graph forms}~\cite{DHoker:2016mwo,DHoker:2016quv}, where an $SL(2,\ints)$-invariant or covariant function is associated with a certain graph that is to be thought of as describing a Feynman diagram on the toroidal world-sheet. The modular function is then obtained from the graph by Feynman-type rules and the resulting functions quickly go beyond the usual types of holomorphic or non-holomorphic modular forms when considering complicated Feynman diagrams. Understanding them is crucial for exploring the structure of the low-energy expansion of string theory at one-loop order, see~\cite{Green:1999pv,Green:2008uj,DHoker:2015wxz,Basu:2016kli,Basu:2016fpd,Kleinschmidt:2017ege,Basu:2017nhs,Basu:2017zvt,Gerken:2018jrq,Gerken:2018zcy} for further work on this topic for type II and heterotic strings.

In the second instance, the group $SL(2,\ints)$ is interpreted as the non-perturbative U-duality group of ten-dimensional type IIB string theory~\cite{Hull:1994ys}. Now, $SL(2,\ints)$ acts on the axio-dilaton of the string that includes the string coupling $\gs$ and is thus a non-perturbative symmetry, relating perturbative and non-perturbative effects in $\gs$. 
The symmetry, together with differential constraints from supersymmetry, is powerful and very constraining and can serve to predict effects in string theory and its low-energy effective approximation that are hard to compute otherwise~\cite{Green:1997tv,Kiritsis:1997em,Pioline:1998mn,Green:1998by,Obers:1999um,Green:2005ba,Green:2014yxa}, if one can determine the exact function invariant under $SL(2,\ints)$, an increasingly difficult task as one progresses in the low-energy approximation, \textit{e.g.}, higher curvature corrections to the four-graviton sector of the form $D^{2k} R^4$ with increasing $k$.

\subsection{Inhomogeneous equations and Poincar\'e series}

A common feature of both instances is that the $SL(2,\ints)$-invariant functions that arise generically satisfy inhomogeneous Laplace equations. Denoting the variable on which $SL(2,\ints)$ acts by $z=x+i y\in \UHP$ for both cases, this differential equation is of the form
\begin{align}
\label{eq:inhLap}
\left( \Delta - s(s+1) \right) f(z) = R(z)\,,
\end{align}
where $\Delta= y^2 \left(\partial_x^2 + \partial_y^2\right)$ is the $SL(2)$-invariant scalar Laplacian and $R(z)$ an $SL(2,\ints)$-invariant right-hand side. In the first instance (modular graph functions), $f(z)$ has a known representation as a multi-lattice sum, whereas in the second instance $f(z)$ is in general unknown. But even in the multi-lattice sum case, it is often not obvious how to extract different explicit forms of the modular graph function, such as the Fourier expansion that contains (elliptic) single-valued multi-zeta values~\cite{Zerbini:2015rss,DHoker:2017zhq,Broedel:2018izr}. 

The aim of this paper is to provide tools for analysing $f(z)$ in the case when $R(z)$ can be represented as a convergent \textit{Poincar\'e series}
\begin{align}
\label{eq:PSrhs}
R(z) = \sum_{\gamma \in B(\ints) \backslash SL(2,\ints)} \rho(\gamma z)
\end{align}
with the standard $SL(2,\ints)$ action
\begin{align}
\gamma = \begin{pmatrix}a &b\\c&d \end{pmatrix}\in SL(2,\ints) \quad \Rightarrow \quad
\gamma z= \frac{az+b}{cz+d}\,,
\end{align}
and where the Borel subgroup
\begin{align}
B(\ints) = \left\{ \pm \begin{pmatrix}  1 &m \\ 0 &1\end{pmatrix} \,\middle|\, m\in\ints \right\} \subset SL(2,\ints)
\end{align}
thus acts by translations $z\mapsto z+ m$. The quotient by $B(\ints)$ in the Poincar\'e sum~\eqref{eq:PSrhs} indicates that the function $\rho(z)$ is periodic, $\rho(z+m)=\rho(z)$ for all $m\in\ints$, and is necessary in order to avoid divergences. 

A representation of the form~\eqref{eq:PSrhs} for $R(z)$ is easy to obtain for the case of the simplest modular graph functions and for the $D^6R^4$ correction coefficient as it is known in this case that $R(z)$ is a polynomial (of second order) in non-holomorphic Eisenstein series
\begin{align}
\label{eq:Eis}
E_s(z) &= \sum_{\gamma \in B(\ints)\backslash SL(2,\ints)} \left[\Im(\gamma z)\right]^s\nn\\
& = y^s + \frac{\xi(2s-1)}{\xi(2s)} y^{1-s} + \frac{2}{\xi(2s)} y^{1/2} \sum_{n\neq 0} |n|^{s-1/2} \sigma_{1-2s}(n) K_{s-1/2} (2\pi | n| y) e^{2\pi i n x}\,,
\end{align}
whose Poincar\'e series is absolutely convergent for $\Re(s)>1$. In the second line, we have given the Fourier expansion of $E_s(z)$ in terms of the completed Riemann zeta function $\xi(k)\!=\! \pi^{-k/2} \Gamma(k/2) \zeta(k)$, the divisor sum $\sigma_k(n) = \sum_{d | n} d^k$ and the modified Bessel function $K_s(y)$. The Eisenstein series $E_s(z)$ has a standard analytic continuation to $\Re(s)<1$ (obtainable from~\eqref{eq:Eis}) and also satisfies the functional relation $\xi(2s)E_s(z)= \xi(2(1-s)) E_{1-s}(z)$. The continuation implies in particular that $E_0(z) =1$.

The strategy proposed in~\cite{Ahlen:2018wng} (see also~\cite{DHoker:2015gmr,DHoker:2019txf}) is to represent the function $f(z)$ in~\eqref{eq:inhLap} also as a Poincar\'e series
\begin{align}
\label{eq:PSf}
f(z) = \sum_{\gamma \in B(\ints) \backslash SL(2,\ints)} \sigma(\gamma z)
\end{align}
with a periodic `seed function' $\sigma(z)=\sigma(z+m)$ for all $m\in\ints$. If this sum is absolutely convergent, one may attempt to solve the equation
\begin{align}
\label{eq:inhLap2}
\left( \Delta - s(s+1) \right) \sigma(z) = \rho(z)
\end{align}
instead of~\eqref{eq:inhLap} due to the $SL(2,\ints)$-invariance of the Laplacian. The complexity of this equation is typically less than that of the original equation~\eqref{eq:inhLap}. The price to pay for this is that one has to analyse in more detail the convergence of~\eqref{eq:PSf}, that one has to provide appropriate boundary conditions for~\eqref{eq:inhLap2} and lastly that one only has indirect information about $f(z)$ through~\eqref{eq:PSf} even if $\sigma(z)$ is completely known.

\subsection{Outline}

In this paper, we shall study how to extract information about the Fourier expansion of $f(z)$ from that of $\sigma(z)$ and will address different subtleties that arise. We focus mainly on the zero Fourier mode of $f(z)$ and study its perturbative part (power series in $y$) and non-perturbative part (power series in $e^{-2\pi y}$). In the case of $SL(2,\ints)$ U-duality, these terms are interpreted as string perturbative and instanton/anti-instanton corrections to the higher-derivative correction term. In the case of modular graph functions, they encode the asymptotic behaviour of the one-loop scattering amplitude as the torus world-sheet degenerates to a very thin torus of large diameter. 

In Section~\ref{sec:strat}, we review how to generally relate the Fourier expansion of $f(z)$ to its Poincar\'e seed $\sigma(z)$ and a certain important proposition of Zagier's that can be used to extract from this the asymptotic expansion of the zero mode of $f(z)$. This procedure will be carried out for a general class of seeds $\sigma(z)$ in Section~\ref{sec:General} and the subsequent sections contain several examples of modular graph functions and the $D^6R^4$ higher-derivative correction that all can be mapped back to this general class. Two appendices contain complementary technical details for some of the calculations carried out in the main body of the paper.

\section{General strategy}
\label{sec:strat}

It is well-known how to relate the Fourier expansion of $f(z)$ to that of $\sigma(z)$ if the two are related through the convergent Poincar\'e series relation~\eqref{eq:PSf}, see for instance~\cite{Iwaniec,Fleig:2015vky} or the brief review in Appendix \ref{app:Kloost}. If the Fourier expansions of $f(z)$ and $\sigma(z)$ are given by
\begin{align}
f(z) = \sum_{n\in \ints} a_n(y) e^{2\pi i n x}\,,
\quad
\sigma(z) = \sum_{n\in \ints} c_n(y) e^{2\pi i n x}\,,
\end{align}
then one has
\begin{align}\label{eq:NZmodes}
a_n(y) = c_n(y) + \sum_{c>0} \sum_{m\in\ints} S(m,n;c) \int_\reals e^{-2\pi i n \omega - 2\pi i m \frac{\omega}{c^2(y^2+\omega^2)}} c_m\left( \frac{y}{c^2(y^2 + \omega^2)} \right) d\omega\,.
\end{align}
Here, $S(m,n;c)$ denotes in general a Kloosterman sum
\begin{align}
S(m,n;c) = \sum_{q \in (\ints/c\ints)^\times} e^{2\pi i (mq + nq^{-1})/c}\,,
\end{align}
a finite sum over all $0\leq q < c$ that are co-prime to $c$ such that $q$ is a multiplicatively invertible element of $\ints/c\ints$ as indicated in the sum. 

Our main interest at present lies with the zero mode $a_0(y)$ and therefore we are faced with making sense of
\begin{align}
\label{eq:zeroint}
a_0(y) &= c_0(y) + \sum_{c>0} \sum_{m\in\ints} \sum_{q\in (\ints/c\ints)^\times} e^{2\pi i m q/c} \int_\reals e^{- 2\pi i m \frac{\omega}{c^2(y^2+\omega^2)}} c_m\left( \frac{y}{c^2(y^2 + \omega^2)} \right) d\omega\nn\\
&= c_0(y)+ y \sum_{c>0}  \sum_{m\in\ints} \sum_{q\in (\ints/c\ints)^\times} e^{2\pi i m q/c} \int_\reals e^{- 2\pi  m \frac{i t}{y c^2(1+t^2)}} c_m\left( \frac{y^{-1}}{c^2(1 + t^2)} \right) dt\nn\\
&= c_0(y)+y \sum_{c>0} \sum_{q\in (\ints/c\ints)^\times}  \int_\reals c_0\left( \frac{y^{-1}}{c^2(1 + t^2)} \right) dt + I \,,
\end{align}
for a given set of Fourier modes $c_m(y)$ of a seed $\sigma(z)$. Here, we have introduced the separate notation
\begin{align}
I = y \sum_{c>0}  \sum_{m\neq 0} \sum_{q\in (\ints/c\ints)^\times} e^{2\pi i m q/c} \int_\reals e^{- 2\pi  m \frac{i t}{y c^2(1+t^2)}} c_m\left( \frac{y^{-1}}{c^2(1 + t^2)} \right) dt
\end{align}
for the contribution from the non-zero modes $c_m$ with $m\neq 0$ to $a_0$.

We are interested in the asymptotic expansion of this expression around $y\to\infty$. We see that the exponent in the integral contains the combination $my^{-1}$ and also the typical seed Fourier modes contain $m$ accompanied by $y^{-1}$ after some manipulations, so that we are faced with expanding a sum of a function evaluated at multiples of its argument asymptotically.
Zagier has proved a very useful result for this situation~\cite{ZagierApp}. Consider a smooth function $\varphi(t)$ for $t >0$ such that itself and all its derivatives are of rapid decay at infinity. Assume also that around the origin one has the asymptotic series $\varphi(t)\sim \sum_{n\geq 0}  b_n t^n$. Then 
\begin{align}
\label{eq:Zagexp}
\sum_{m \geq 0} \varphi( (m+a) t) \sim \frac{I_\varphi}{t} + \sum_{n\geq 0} b_n \zeta(-n,a) t^n
\end{align}
as asymptotic expansion around $t=0$ for the periodic sum with $a>0$. Here, $\zeta(-n,a)$ is the Hurwitz zeta function\footnote{We note for reference that the Hurwitz zeta function is related to the Bernoulli polynomials by the  relation $\zeta(1-k,a) = - \frac{B_{k}(a)}{k}$ for positive $k$. } and 
\begin{align}
I_\varphi = \int_0^\infty \varphi(t) dt\,.
\end{align}

Let us briefly indicate where the terms come from. Plugging in na\"ively the expansion of $\varphi(t)$ leads to~\cite{ZagierApp}
\begin{align}
\sum_{m\geq 0} \varphi((m+a)t) &\underset{\text{na\"ive}}{\sim} \sum_{m\geq 0} \sum_{n \geq 0} b_n (m+a)^n t^n = \sum_{n\geq 0} b_n \left( \sum_{m\geq 0} (m+a)^n\right) t^n
= \sum_{n\geq 0} b_n \zeta(-n,a) t^n\,.
\end{align}
The calculation above is formal since the two sums cannot be interchanged: the $m$-sum is divergent and we have used the Hurwitz zeta function as analytic continuation. 

The other term in~\eqref{eq:Zagexp} is what Zagier calls the Riemann term and can be understood by viewing the sum as an approximation to the Riemann integral for small $t$, where $\frac{1}{t}$ is the length of the interval.\footnote{This is the term that was missed in~\cite{Ahlen:2018wng} where only the analytically continued $\zeta$ terms were found. This explains why~\cite{Ahlen:2018wng} was off at one specific power of the asymptotic variable.}

Zagier has also proved extensions of~\eqref{eq:Zagexp} for the case when $\varphi(t)$ is not $C^\infty$ at the origin but includes terms of the form $t^s \log t$ or $t^{-s}$ for $s> 0$, see~\cite{ZagierApp}.

\section{Asymptotic expansion for general seed}
\label{sec:General}

We will now present the details of the strategy outlined in the previous section and we consider the asymptotic expansion of the Poincar\'e series associated to a very general type of seed function. In particular suppose the non-zero Fourier mode of the seed is of the form
\begin{align}
\label{eq:FMgen}
c_m(y) = \sigma_a(|m|) (4\pi|m| )^b y^r e^{-2\pi|m| y}
\end{align}
for parameters $a$, $b$ and $r$. This is true, up to an overall constant, for all the modular graph functions and all other concrete examples we shall present later are (possibly infinite) combinations of such terms.

Plugging~\eqref{eq:FMgen} into the relevant part of the Fourier mode of the Poincar\'e sum~\eqref{eq:zeroint} leads to a contribution from the non-zero modes of the form
\begin{align}
\label{eq:Igen}
I &=  y \sum_{c>0} \sum_{m\neq 0} \sum_{q\in(\ints/c\ints)^\times} \theta^{mq} \int_\reals e^{-2\pi m \frac{it}{yc^2(1+t^2)}} \sigma_a(|m|) \frac{(4\pi |m|)^b}{(yc^2(1+t^2))^{r}} e^{-2\pi |m| \frac{1}{yc^2(1+t^2)}} dt\nn\\
&= 2y^{1+b-r} \sum_{c>0} c^{-2r+2b}\sum_{m>  0} \sum_{q\in(\ints/c\ints)^\times} \theta^{mq} \sigma_a(m) \left(\frac{4\pi m}{yc^2}\right)^b\int_\reals e^{-2\pi m \frac{1+ it}{yc^2(1+t^2)}} \frac{1}{(1+t^2)^{r}} dt \\
&= \frac{2^{3-2r+2b}\pi y^{1+b-r}}{\Gamma(r)} \!\sum_{c>0} c^{-2r+2b}  \hspace{-4mm}\sum_{q\in(\ints/c\ints)^\times} \sum_{m> 0}\!\theta^{mq} \sigma_a(m) \left(\frac{\pi m}{yc^2}\right)^b \!\!\sum_{k\geq 0} \frac{(-\pi my^{-1} c^{-2})^k}{k!} \frac{\Gamma(2r+k-1)}{\Gamma(r+k)}\,,\nn
\end{align}
when using~\eqref{eq:auxint1} for the integral and having defined $\theta=e^{2\pi i /c}$. The parameter $r$ must have $\Re(r)>1/2$ for the integral to converge so we will assume it in what follows. For positive integer $r$, the quotient of $\Gamma$ functions becomes a polynomial in $k$ such that one could write this as a polynomial in times $e^{-\pi m/yc^2}$. In general the sum over $k$ produces the hypergeometric function ${}_1F_1(2r-1,r;-\pi m/yc^2)$, but we shall leave it as a sum.

In order to determine the asymptotic behaviour of this expression we analyse first the sum over $m$ that can be written as
\begin{align}
\label{eq:genZ}
&\quad \quad\sum_{m> 0}\theta^{mq} \sigma_a(m) \left(\frac{\pi m}{yc^2}\right)^b \sum_{k\geq 0} \frac{(-\pi my^{-1} c^{-2})^k}{k!} \frac{\Gamma(2r+k-1)}{\Gamma(r+k)}\nn\\
&= \sum_{h=1}^c \sum_{m \geq 0} \sum_{n>0} \theta^{nhq} n^{a+b} ((m+\tilde{h})t)^b \sum_{k\geq 0} \frac{(-(m+\tilde{h})t)^k}{k!} n^k \frac{\Gamma(2r+k-1)}{\Gamma(r+k)}
\end{align}
by writing out the divisor and grouping terms in additive classes modulo $c$, and where we have introduced $\tilde{h}=\frac{h}{c}$ and $t=\frac{\pi}{yc}$.

The sum over $m$ in~\eqref{eq:genZ} is of the general form studied in~\cite{ZagierApp} and thus amenable to formula~\eqref{eq:Zagexp}. We assume $b>-1$ and not an integer at first, other cases can be obtained from the final result by analytic continuation. As~\eqref{eq:genZ} is already expanded in powers of $t$, we can immediately write down the asymptotic expansion for fixed $h$ as
\begin{align}
\label{eq:genZ2}
&\quad\quad \sum_{m \geq 0} \sum_{n>0} \theta^{nhq} n^{a+b} ((m+\tilde{h})t)^b \sum_{k\geq 0} \frac{(-(m+\tilde{h})t)^k}{k!} n^k \frac{\Gamma(2r+k-1)}{\Gamma(r+k)} \nn\\
&\sim \frac{I_h}{t} + \delta_{h,c} \,\zeta(a+1) \frac{\Gamma(a+b+1)\Gamma(2r-a-b-2)}{\Gamma(r-a-b-1)}t^{-a-1}\nn\\
&\quad + t^b \sum_{n\geq 0} \frac{(-t)^n}{n!} \zeta\big(-b-n,\frac{h}{c}\big) \frac{\Gamma(2r+n-1)}{\Gamma(r+n)}\Li_{-a-b-n}(\theta^{hq})\,,
\end{align}
where the extra term for $h\equiv c\, (\text{mod}\,c)$ is the `Riemann term' in the asymptotic expansion of the sum over $n$  while the `Riemann integral term' $I_h$ in the sum over $m$ is given for all $h$ by
\begin{align}
I_h = \int_0^\infty \sum_{n>0} \theta^{nhq} n^{a+b} t^b \sum_{k\geq 0} \frac{(-nt)^k}{k!} \frac{\Gamma(2r+k-1)}{\Gamma(r+k)} dt = \frac{\Gamma(b+1)\Gamma(2r-b-2)}{\Gamma(r-b-1)} \Li_{1-a}(\theta^{hq})
\end{align}
when using~\eqref{eq:auxint2}. Note also that the quotient of gamma functions in~\eqref{eq:genZ2} is again a polynomial in $n$ for positive integral $r$.

Summing this over $h$, $q$ and $c$ can be done with the help of formulas~\eqref{eq:zetaPL2} and~\eqref{eq:PL1} to obtain the following expression for $I$ of~\eqref{eq:Igen}:
\begin{align}
\label{eq:IgenAsy}
I&\sim \frac{2^{3-2r+2b}\pi y^{1+b-r}}{\Gamma(r)} \Bigg[ \frac{y}{\pi} \frac{\Gamma(b+1)\Gamma(2r-b-2)}{\Gamma(r-b-1)} \frac{\zeta(2r-a-2b-2)\zeta(1-a)}{\zeta(2r-a-2b-1)} \nn\\
&\hspace{2mm} + \left(\frac{y}{\pi}\right)^{a+1} \frac{\Gamma(a+b+1)\Gamma(2r-a-b-2)}{\Gamma(r-a-b-1)}\frac{\zeta(2r-a-2b-2)\zeta(a+1)}{\zeta(2r-a-2b-1)} \nn\\
&\hspace{2mm} +\left(\frac{\pi}{y}\right)^b 
\sum_{n \geq 0} \left(\frac{-\pi}{y}\right)^{n} \frac{\Gamma(2r+n-1)}{n! \cdot \Gamma(r+n)} \nn\\
&\hspace{30mm}\times \frac{\zeta(-b-n)\zeta(-a-b-n)\zeta(2r-a-b+n-1)\zeta(2r-b+n-1)}{\zeta(2r+2n)\zeta(2r-a-2b-1)} \bigg]\nn\\
&\equiv I(a,b,r)\,.
\end{align}
Here, we have set $\Li_s(1)=\zeta(s)$ by analytic continuation and introduced a short-hand notation for terms of this type. 
The powers of $y$ appearing in the above expression are $y^{2+b-r}$, $y^{2+a+b-r}$ and then $y^{1-r-n}$ for $n\geq 0$.  

We also note that the quotient of zeta functions appearing can be rewritten as a Dirichlet series of two divisor sums as
\begin{align}
&\label{eq:RatioZeta}\hspace{10mm}\frac{\zeta(-b-n)\zeta(-a-b-n)\zeta(2r-a-b+n-1)\zeta(2r-b+n-1)}{\zeta(2r+2n)}\nn\\
&=\frac{ 4\sin\left(\frac{\pi(b+n)}2\right)\sin\left(\frac{\pi(a+b+n)}2\right) \Gamma(1+b+n)\Gamma(1+a+b+n)}{(2\pi)^{a+2b+2n+2}}\nn\\
&\hspace{15mm} \times
\sum_{m>0} \sigma_{a}(m) \sigma_{a+2b+2-2r}(m) m^{-1-a-b-n}\,
\end{align}
where we have used the functional equation for the Riemann zeta function and an identity of Ramanujan's. The only $n$-dependence is in the exponent of the new summation variable $m$, but with this rewriting it appears manifest that for general parameters $a,b,$ and $r$, the sum over $n$ is a factorially divergent asymptotic series. 

We will now provide some concrete examples and show how the procedure described above allows us to reproduce the perturbative expansion of the zero Fourier mode for certain Poincar\'e sums and how it can also be used to retrieve the non-perturbative, exponentially suppressed terms.

\section{Modular graph functions}

In this section, we apply the method outlined in Section~\ref{sec:strat} to modular graph functions by making use of expression~\eqref{eq:IgenAsy} derived in the previous section. Rather than giving a general analysis, we present two exemplary cases that highlight also how to deal with certain analytical continuations and singularities.

\subsection{The \texorpdfstring{$C_{3,1,1}$}{C311} modular graph function}
\label{sec:311}

We consider first the example of the modular graph function $C_{3,1,1}(z)$ that is given explicitly by the double-lattice sum~\cite{DHoker:2015gmr}
\begin{align}
C_{3,1,1}(z) = \sum_{\substack{(m_1,n_1), (m_2,n_2) \in \ints^2 \\[1mm] (m_i,n_i)\neq (0,0)\\[1mm](m_1+m_2,n_1+n_2)\neq (0,0)}} \frac{y^5}{\pi^5 |m_1z+n_1|^6 |m_2z+n_2|^2 |(m_1+m_2)z + (n_1+n_2)|^2}\,,
\end{align}
and its zero mode has the known terms~\cite{Green:2008uj,DHoker:2015gmr}
\begin{align}
\label{eq:a0311}
a_0(y) = \frac{2 \pi^5}{155\, 925} y^5 + \frac{2\pi^2\zeta(3)}{945} y^2 -\frac{\zeta(5)}{180} + \frac{7\zeta(7)}{16\pi^2} y^{-2} -\frac{\zeta(3)\zeta(5)}{2\pi^3} y^{-3} + \frac{43\zeta(9)}{64\pi^4} y^{-4} + O(e^{-4\pi |n| y})\,.
\end{align}
This result was derived by studying the zero mode of the differential equation satisfied by $C_{3,1,1}(z)$ directly, with appropriate boundary conditions.  We shall rederive this result from the Poincar\'e series method outlined above.

The function $f(z)=C_{3,1,1}(z)$ satisfies the differential equation~\cite{DHoker:2015gmr}\footnote{Note that the Eisenstein series in~\cite{DHoker:2015gmr} are normalised differently from those in~\eqref{eq:Eis}.}
\begin{align}
\label{eq:Lap311}
\left( \Delta - 6 \right) f(z) =  \frac{172 \pi^5}{467\, 775} E_5(z) - \frac{8\pi^5}{42\,525} E_2(z)E_3(z) + \frac{\zeta(5)}{10}\,.
\end{align}
Using the Poincar\'e series representation~\eqref{eq:Eis} for $E_3$ and $E_5$ we deduce that the seed function $\sigma(z)$ for $f(z)$ has to satisfy the differential equation
\begin{align}
\label{eq:Ls311}
\left( \Delta - 6 \right) \sigma(z) = \frac{172 \pi^5}{467\, 775} y^5  - \frac{8\pi^5}{42\,525} y^3 E_2(z)+ \frac{\zeta(5)}{10} y^\epsilon\,.
\end{align}
We have introduced a regulating $y^\epsilon$ for the constant term $\zeta(5)/10$ in~\eqref{eq:Lap311}, keeping in mind that $\lim_{\epsilon\to 0} E_\epsilon(z) = 1$ is constant by the standard analytic continuation of the convergent Eisenstein series and this shows how to deal with constant sources in the original differential equation.

The differential equation~\eqref{eq:Ls311} can be solved for the Fourier modes $c_m(y)$ of the seed $\sigma(z)=\sum_{n\in\ints} c_n(y) e^{2\pi i n x}$ by~\cite{Ahlen:2018wng}
\begin{align}
\label{eq:sol311}
c_0 (y) &= \frac{2\pi^5}{155\, 925} y^5 + \frac{2\pi^2 \zeta(3)}{945} y^2 + \frac{\zeta(5)}{10(\epsilon(\epsilon-1)-6)} y^\epsilon \,,\nn\\
c_m(y) &= \frac{2\pi^2}{945} \sigma_{-3}(\vert m\vert) y^2 e^{-2\pi |m| y}\,, \hspace{40mm} (m\neq 0)\,.
\end{align}
Here, we have used the Fourier expansion of the Eisenstein series from~\eqref{eq:Eis} and the simple form of $c_m(y)$ is tied to the fact that $K_{3/2}$ appearing in $E_2$ has an exact expansion around $y\to \infty$. The particular solution of the Laplace equation (\ref{eq:Ls311}) above has been fixed by requiring the correct asymptotic behaviour fixed uniquely from the behaviour of the right-hand side.

\subsubsection{Perturbative contributions to the zero mode}

Note that the seed function~\eqref{eq:sol311} is precisely of the type~\eqref{eq:FMgen} described in the previous section when we substitute $a=-3,\, b=0,\, r=2$. As it will be shortly clear it is better to keep $b$ as a regulator and send it to zero only at the very end, this is by no mean necessary but it allows us to have a uniform description given by the asymptotic expansion of seeds of the type (\ref{eq:FMgen}). (Otherwise, one could use the extension of Zagier's method to include $\log$-terms.)

The solution~\eqref{eq:sol311} can now be substituted into~\eqref{eq:zeroint} to obtain the zero mode of $C_{3,1,1}(z)= \sum_{n\in\ints} a_n(y) e^{2\pi i n x}$ 
and the contribution $I$ from all $c_m(y)$ with $m\neq 0$ to $a_0(y)$ is precisely captured by the general formula (\ref{eq:IgenAsy}) specialized to the present case
\begin{align}
\label{eq:I311}
I &\sim \lim_{b \to 0} \frac{2\pi^2}{945}  I(-3,\,b,\,2)\,,\\
&\nn = \lim_{b \to 0} \frac{4^b \pi^3}{945 y^{1-b}}\Bigg[ \frac{ y}{\pi} \frac{\Gamma(1+b)\Gamma(2-b)}{\Gamma(1-b)} \frac{\zeta(5-2b)\zeta(4)}{\zeta(6-2b)}+\left(\frac{\pi}{y}\right)^{b}\sum_{n\geq0} \left(\frac{-\pi}{y}\right)^n\frac{(n+2)}{n!}   \\
&\nn\hspace{35mm} \times \frac{\zeta(-n - b)\zeta(3 - n - b) \zeta(
  3 + n - b) \zeta(6 + n - b)}{
\zeta(4 + 2 n) \zeta(6 - 2 b)} \Bigg]\,,
\end{align}
here the second term of (\ref{eq:IgenAsy}) is absent since it is proportional to $\zeta(a+1)$ that in the present case is $\zeta(-2)=0$.

It is simple to see that $b$ serves as a regulator only for the $n=2$ term in the infinite series by producing a finite limit for the $\zeta(-2-b) \zeta(1-b)$ term and can be set to zero in all the remaining terms. Furthermore this asymptotic series does actually truncate when $b \to 0$ due to the presence of the first two zetas $\zeta(-n - b)\zeta(3 - n - b) \to 0$ when $n\geq4$ since either one of the two zetas will be evaluated at a negative even integer, hence vanishing.
This opposite parity, crucial for the truncation of the asymptotic series, can be traced back to the odd index of the divisor function $\sigma_{-3}(\vert m \vert)$, appearing in the seed function (\ref{eq:sol311}). For general index (in particular even) divisors the asymptotic expansion (\ref{eq:IgenAsy}) will not truncate.

With these considerations in mind it is fairly simple to take the limit for $b \to 0$ in (\ref{eq:I311}) obtaining
\begin{equation}
\label{eq:I311asy}
I \sim \frac{\zeta(5)}{90} - \frac{2\zeta(3)^2}{21 \pi y} + \frac{7\zeta(7)}{16\pi^2 y^2} - \frac{\zeta(3)\zeta(5)}{2\pi^3y^3} + \frac{11 \zeta(9)}{32\pi^4 y^4}\,.
\end{equation}

As the zero mode of $C_{3,1,1}(z)$ is given according to~\eqref{eq:zeroint} by
\begin{align}
a_0(y) = c_0(y) + y \sum_{c>0} \sum_{q\in(\ints/c\ints)^\times} \int_\reals c_0\left(\frac{y^{-1}}{c^2(1+t^2)}\right)dt +I \,,
\end{align}
and we have computed the asymptotic expansion of $I$, it remains to determine the contributions from $c_0(y)$. These are with~\eqref{eq:sol311} given by\footnote{The Euler totient function $\phi(c)$ gives the cardinality of $(\ints/c\ints)^\times$ and has Dirichlet series $\sum_{c>0} \phi(c) c^{-s}  = \frac{\zeta(s-1)}{\zeta(s)}$.}
\begin{align}
\label{eq:c0311}
c_0(y) + y \sum_{c>0} \phi(c) \int_\reals c_0\left(\frac{y^{-1}}{c^2(1+t^2)}\right)dt 
&\underset{\epsilon\to 0}{\to} 
\frac{2\pi^5 y^5}{155\, 925} + \frac{2\pi^2 \zeta(3) y^2}{945}  - \frac{\zeta(5)}{60} + \frac{2\zeta(3)^2}{21\pi y} + \frac{21\zeta(9)}{64\pi^4y^4} \,.
\end{align}
Compared to the asymptotic calculation of $I$, this is an exact result and we have taken the limit of $\epsilon\to 0$ after performing the integral of $c_0$.

Combining~\eqref{eq:I311asy} and~\eqref{eq:c0311} leads to the full asymptotic zero mode perturbative expansion of $C_{3,1,1}(z)$ as already presented in~\eqref{eq:a0311}.
As we will see shortly see it will also be possible to extract from equation (\ref{eq:I311asy}) the complete non-perturbative completion of the zero mode which is entirely captured by the perturbative data.

\subsubsection{Non-perturbative terms}
\label{sec:NP}

We have just reconstructed the full perturbative expansion of the zero-mode of the $C_{3,1,1}(z)$ modular graph function, however it is simple to see from the partial differential equation (\ref{eq:Lap311}) that due to the Eisenstein series (\ref{eq:Eis}), this zero-mode will need to receive infinitely many non-perturbative corrections of the the form $e^{- 4\pi m y}$ with $m\in \mathbb{N}$.

Given an asymptotic power series, a standard approach to reconstruct the full non-perturbative contributions out of the perturbative data is given by resurgent analysis and Borel--Ecalle resummation \cite{Ecalle:1981} (see also \cite{delabaere1999resurgent}), but unfortunately this is not directly amenable to the present case due to the truncation of the perturbative series (\ref{eq:a0311}) to a finite number of terms. Here, we only need to consider the contribution from $I$ in~\eqref{eq:I311} since the zero mode $c_0(y)$ does not produce an asymptotic tail but only gives rise to a simple Laurent polynomial in $y$ as presented in equation (\ref{eq:c0311}).

However, thanks to our analysis of the asymptotic expansion of this Poincar\'e series via the seed function presented in equation (\ref{eq:I311asy}) we will see that we can reconstruct the complete non-perturbative transseries expansion for (\ref{eq:a0311}) entirely out of the purely perturbative data (\ref{eq:I311asy}) in a beautiful example of Cheshire-cat resurgence \cite{Dunne:2016jsr,Kozcaz:2016wvy,Dorigoni:2017smz}.

Our starting point is the asymptotic expansion (\ref{eq:I311}) before taking the $b \to 0$ limit 
\begin{align}
\label{eq:I311asyeps}
I(b) =  I_{\text{pert}}(b)+ \frac{4^\epsilon \pi^{3+b}}{945 y}\sum_{n\geq4} \left(\frac{-\pi}{y}\right)^n\frac{(n+2)}{n!}  \frac{\zeta(-n - b)\zeta(3 - n - b) \zeta( 3 + n -b) \zeta(6 + n - b)}{\zeta(4 + 2 n) \zeta(6 - 2 b)}\,,
\end{align}
where we split the infinite asymptotic series into the sum of a piece
\begin{align}
I_{\text{pert}}(b) &=  \frac{4^b \pi^2}{945 y^{-b}} \frac{\Gamma(1+b)\Gamma(2-b)}{\Gamma(1-b)} \frac{\zeta(5-2b)\zeta(4)}{\zeta(6-2b)} \\
&\hspace{10mm}+\frac{4^b \pi^{3+b}}{945 y}\sum_{n=0}^3 \left(\frac{-\pi}{y}\right)^n\frac{(n+2)}{n!}  \frac{\zeta(-n - b)\zeta(3 - n - b) \zeta(  3 + n - b) \zeta(6 + n -b)}{\zeta(4 + 2 n) \zeta(6 - 2 b)}\nn
\end{align}
with non-vanishing $b\to 0$ limit, reproducing precisely the perturbative expansion (\ref{eq:I311asy}), and an asymptotic tail that vanishes when $b \to0$.

To make this manifest we can rewrite the asymptotic tail using Riemann's functional equation and shift $n\to n+4$ to obtain
\begin{align}
 \label{eq:C311As} 
I(b) &= I_{\text{pert}}(b) +\frac{16 \pi^{5-b}}{945 \,\zeta(6-2b)} \sin(\pi b) \sum_{n\geq 0}  ( 4\pi y)^{-n-5} \frac{(n+6) \Gamma( n +2+ b)\Gamma(n+5 + b)}{(n+4)!} \nn\\
&\hspace{30mm}\times \frac{\zeta(5+n+b)\zeta(2+n+b)\zeta(7+n-b) \zeta(10+n-b)}{\zeta(2n+12)}\,.
\end{align}

As anticipated the tail is a Gevrey-1 asymptotic series, \textit{i.e.}, growing like $n!$, regular in the $b \to0$ limit and multiplied by $\sin(\pi b)$ that also vanishes manifestly in the same limit.
Our strategy will be now to keep this vanishing $\sin(\pi b)$ while setting $b =0$ in all the remaining regular terms\footnote{This is not necessary and one can repeat this analysis while keeping $b\neq0$ in all terms, the expression will just be slightly more involved but the final result for the non-perturbative corrections will not change when we take $b\to 0$.} and then try to perform a Borel--Ecalle resummation (see for example \cite{delabaere1999resurgent}) of the asymptotic tail
\begin{equation}\label{eq:IasySeries}
I_{\text{asy}}(b)= \frac{16}{\pi} \sin(\pi b) \sum_{n\geq 0} ( 4\pi y)^{-n-5} (6+n) \Gamma(n+2) \frac{\zeta(2+n)\zeta(5+n)\zeta(7+n) \zeta(10+n)}{\zeta(2n+12)}\,.
\end{equation}

To proceed with this strategy we would need first to find a measure $d\mu(t)$, say for $ t\in \mathbb{R}^+$, whose momenta
\begin{equation}
\int_0^\infty t^{n+5} \, d\mu(t) =d_n\label{eq:Momenta}
\end{equation} 
have the same asymptotic growth as the coefficients $c_n=(6+n) \Gamma(n+2) \frac{\zeta(2+n)\zeta(5+n)\zeta(7+n) \zeta(10+n)}{\zeta(2n+12)}$ of equation (\ref{eq:IasySeries}). More precisely we require the modified Borel transform
\begin{equation}
B(t)  = \sum_{n\geq0}\frac{c_n}{d_n} t^{n+5}\,,\label{eq:BorelGenerel}
\end{equation}
to have finite radius of convergence, thus defining a germ of analytic functions at the origin $t=0$.
If this happens we can commute the formal series in equation (\ref{eq:IasySeries}) with the above integral to define a possible resummation of the original asymptotic series via
\begin{equation}
 \frac{16}{\pi} \sin(\pi b) \int_0^\infty \sum_{n\geq0}\frac{c_n}{d_n}\left(\frac{t}{4\pi y}\right) ^{n+5} \, d\mu(t)=\frac{16}{\pi} \sin(\pi b) \int_0^{\infty} B\left(\frac{t}{4\pi y}\right) \, d\mu(t)\,.
\end{equation}
Under certain reasonable assumptions (see \cite{Ecalle:1981,delabaere1999resurgent}), this integral defines an analytic function in a certain sector of the complex $y$-plane whose asymptotic expansion for $y\gg1$ coincides with (\ref{eq:IasySeries}).

The usual Borel kernel amounts to considering the simple measure $d\mu(t) = e^{-t}\,t^\alpha\,dt$ so that the momenta (\ref{eq:Momenta}) are simply $d_n = \Gamma(n+4+\alpha)$. In the present case we could consider this measure however we would not be able to compute analytically the standard Borel transform for the given coefficients $c_n$ of equation (\ref{eq:IasySeries}) due to the presence of the ratio of Riemann zetas.
An alternative would be to find a measure whose momenta $d_n$ cancel in the Borel transform not only the factorial growth of the coefficients $c_n$ but also this particular ratio of zetas, so to make the modified Borel transform (\ref{eq:BorelGenerel}) amenable to calculation, however no such measure is known to us.\footnote{
It is interesting to notice that the coefficients of the asymptotic tail are  schematically of the form $\Gamma(n+\alpha) \Pi_{i=1}^4 \zeta(n+\alpha_i)/\zeta(2n+\beta)$. Had there been instead only a single zeta function multiplying the gamma we would have known a measure \cite{Russo:2012kj} (see also \cite{Hatsuda:2015owa}) whose momenta would produce both:
\begin{equation}
\int_0^\infty t^n \frac{dt}{4\sinh^2(t/2)} = \Gamma(n+1)\zeta(n)\,,\nn
\end{equation}
valid for $n\geq 2$,
or more generally we find that the measure $d\mu(t) = {\mbox{Li}}_{\beta-\alpha}(e^{-t})\, t^{\alpha -1} \,dt$ has momenta \begin{equation}
 \int_0^\infty t^n d\mu(t) = \Gamma(n+\alpha)\zeta(n+\beta)\,\nn
\end{equation}
valid for $n>-\alpha$ when $\beta-\alpha>1$, or $n>1-\beta$ when $\beta-\alpha\leq1$. A Borel--Ecalle resummation with this modified kernel is possible and in \cite{Arutyunov:2016etw} it was shown that one can reinterpret an asymptotic series with coefficients of the form $\Gamma(n+\alpha) \zeta(n+\beta)$ as a series with simpler coefficients evaluated at shifted couplings $y\to m y$ with $m\in\mathbb{N}$ for which it is easier to evaluate the full non-perturbative completion. A very similar phenomenon will arise in the present case.}

To proceed, we realise that we can write that particular ratio of zetas, as discussed in equation (\ref{eq:RatioZeta}), in terms of the Dirichlet series
\begin{equation}
\frac{\zeta(2+n)\zeta(5+n)\zeta(7+n) \zeta(10+n)}{\zeta(2n+12)}=\sum_{m>0} \sigma_{-3}(m) \sigma_{-5}(m) m^{-n-2}\,,
\end{equation}
so that we can rewrite (\ref{eq:C311As}) as an infinite series of a very simple asymptotic expansion evaluated at shifted coupling $ y \to m y$ very reminiscent of \cite{Arutyunov:2016etw}:
\begin{equation}
I_{\text{asy}}(b)= \frac{16}{\pi} (4\pi y)^{-3}  \sin(\pi b) \sum_{m>0} \sigma_{-3}(m)\sigma_{-5}(m)\sum_{n\geq0}(4\pi m y)^{-n-2}  (6+n) \Gamma(n+2) \,.\label{eq:C311As1}
\end{equation}

At this point we can just use standard Borel transform to resum the asymptotic tail.
We make use of the known Laplace integral
\begin{equation}
\int_0^\infty e^{- t} (t/z)^{n+1}  dt=  z^{-n-1}\Gamma(n+2)
\end{equation}
to give a well-defined resummation of the formal asymptotic series 
\begin{align}
F(z)=\sum_{n\geq0} z^{-n-2}(6+n) \Gamma(n+2) \mapsto \mathcal{S}_\theta[F](z) &\nn\equiv \int_0^{\infty e^{-i\theta}} e^{- z\,t} \left[ \sum_{n\geq0}t^{n+1} (6+n) \right]  dt \\
&\label{eq:Borel}= \int_0^{\infty e^{-i\theta}} e^{- z\,t}\,\frac{t(6-5t)}{(t-1)^2} dt\,,
\end{align}
where $\theta=\text{arg} \, z$ and $\mathcal{S}_\theta$ denotes what is usually referred to as directional Borel resummation.

We define in this way the resummation of the asymptotic series (\ref{eq:C311As1}):
\begin{equation}
I_{\text{asy}}(b)= \frac{16}{\pi} (4\pi y)^{-3}  \sin(\pi b) \sum_{m>0} \sigma_{-3}(m)\sigma_{-5}(m)\mathcal{S}_\theta [F]( 4\pi m y)\,.\label{eq:C311Res}
\end{equation}
However, we see that since $\text{arg}(4\pi m\,y) = 0$, the relevant region $y>0$, is a Stokes direction, i.e. a singular direction for the Borel transform (\ref{eq:Borel}).
This means that if we define the two lateral resummations
\begin{equation}
\mathcal{S}_{\pm}[F](4 \pi m y) = \lim_{\theta\to0^+} \int_0^{\infty e^{\mp i\theta}} e^{- 4\pi m y t}\frac{t(6-5t)}{(t-1)^2} dt\,,
\end{equation}
we will have a very simple, yet non-zero, discontinuity for (\ref{eq:Borel}) called Stokes automorphism:
\begin{equation}
\mathcal{S}_{-}[F](4 \pi m y)- \mathcal{S}_{+}[F](4 \pi m y) = \oint_{t=1} e^{- 4\pi m y t}\frac{t(6-5t)}{(t-1)^2}dt = -2\pi i  e^{-4\pi m y} (4 + 4\pi m y)\,.
\end{equation}
 
This means that our resummation (\ref{eq:C311Res}) would give rise to ambiguities in defining \textit{the} value for the starting asymptotic series (\ref{eq:C311Res}) when $y>0$, since as just shown we would get two different results by taking the limit $ \text{arg}(4\pi m\,y) \to 0^{\pm}$ in (\ref{eq:C311Res}); furthermore, although the formal expression (\ref{eq:IasySeries}) we started with is manifestly real for $y>0$ neither of the two lateral resummations is. 

To obtain a real and unambiguous resummation for $y>0$ we have to consider a sort of average between the two lateral resummations, usually referred to as median resummation \cite{delabaere1999resurgent},
and define the resummation of (\ref{eq:C311As1})
\begin{equation}
 I_{\text{asy}}(b) =  \frac{16 (4\pi y)^{-3}}{\pi} \! \sum_{m>0} \sigma_{-3}(m)\sigma_{-5}(m)\big[ \sin(\pi b) \mathcal{S}_\pm [F]( 4 \pi m y) -  \pi( \pm i  \sin(\pi b) ) e^{-4\pi m y} (4 + 4\pi m y) \big]\,,\label{eq:C311ResMed}
\end{equation}
where the sign is according to $\text{arg} \,y< 0$ or $\text{arg} \,y> 0$.
We have substracted half of the Stokes automorphism (with sign) from the lateral resummations so that in practice this amounts to use a principal value prescription to compute the lateral resummation (\ref{eq:C311Res}) for $\theta= \text{arg} \,y=0$. Clearly the asymptotic expansion for $y\gg1$ of the above equation coincides with (\ref{eq:C311As1}) since we have only added non-perturbative terms, and precisely thanks to these non-perturbative terms the median resummation produces a real and unambiguous result in the limit $ \text{arg}(4\pi m\,y) \to 0^{\pm}$, i.e. $y>0$.

As we send $b\to 0 $ we see that both the asymptotic tail and the non-perturbative contributions seem to vanish, however, we will make the hypothesis that this is an example of Cheshire-cat resurgence \cite{Dunne:2016jsr,Kozcaz:2016wvy,Dorigoni:2017smz} for which the non-perturbative terms will still be present in this limit despite the vanishing of the asymptotic tail.
We make the assumption that the transseries parameter, i.e. the factor $\sigma= \pm i  \sin(\pi b) $ has also a real part and it exponentiates to $\sigma\to  \exp( \pm i \pi b)$, so that its imaginary part still vanishes in the $b\to0$ limit, while its real part remains and the full transseries becomes 
\begin{align}
 I(b) &\notag = I_{\text{pert}}(b) + \frac{16 }{\pi (4\pi y)^3}\!\!  \sum_{m>0} \!\!\sigma_{-3}(m)\sigma_{-5}(m)\!\left[ \sin(\pi b) \mathcal{S}_\pm [f](4\pi m y)  -  \pi e^{\pm i  \pi b} e^{-4\pi m y} (4 + 4\pi m y) \right]\\
 &\!\!\underset{b\to 0}{\to} I_{\text{pert}}(0) - 16 (4\pi y)^{-3}  \sum_{m>0} \sigma_{-3}(m)\sigma_{-5}(m)  e^{-4\pi m y} (4 + 4\pi m y)\nn\\
 &= I_{\text{pert}}(0) -   \sum_{m>0} \sigma_{-3}(m)\sigma_{-5}(m)  (\pi y)^{-2} m  e^{-4\pi m y} \left(1 + \frac{1}{\pi m y}\right)   \,,
\label{eq:C311ResFinal}
\end{align}
which reproduces precisely the non-perturbative contribution that can be easily extracted from the PDE (\ref{eq:Lap311}) and have been recently discussed in  \cite{DHoker:2019txf}. 
In principle, if we had a PDE formulation for the seed function with $b \neq 0$, it should be possible to extract the complete transseries parameter without having to assume any exponentiation hypothesis, however at the present time this deformed PDE description in $b$ is lacking.

To conclude this section, we want to stress the importance of the result: from a suitable deformation of the seed function we obtain a perturbative expansion of the zero Fourier mode that does not truncate. Using a clever rewriting of the Riemann zeta functions we can write this asymptotic tail as an infinite series of two divisor functions times a simpler asymptotic series evaluated at shifted coupling $4\pi m y$. The Borel--Ecalle resummation of this asymptotic tail forces us to introduce non-perturbative terms of the form $e^{-4\pi m y}$, entirely encoded in the perturbative data, and, under some reasonable assumption, these non-perturbative terms survive even when we remove this $b$ regulator thus giving us the full non-perturbative completion of the zero Fourier mode for the modular graph function.

\subsection{The \texorpdfstring{$C_{2,1,1}$}{C211} modular graph function}
\label{sec:211}

We consider next the example of the modular graph function $C_{2,1,1}(z)$ that is given explicitly by the double-lattice sum
\begin{align}
 C_{2,1,1}(z) = \sum_{\substack{(m_1,n_1), (m_2,n_2) \in \ints^2 \\[1mm] (m_i,n_i)\neq (0,0)\\[1mm](m_1+m_2,n_1+n_2)\neq (0,0)}} \frac{y^4}{\pi^4 |m_1z+n_1|^4 |m_2z+n_2|^2 |(m_1+m_2)z + (n_1+n_2)|^2}\,,
\end{align}
and its zero mode has the known terms~\cite{Green:2008uj,DHoker:2015gmr}
\begin{align}
\label{eq:a0211}
a_0(y) = \frac{2 \pi^4}{14\, 175} y^4 + \frac{\pi \zeta(3)}{45} y +\frac{5 \zeta(5)}{12 \pi}y^{-1} - \frac{\zeta(3)^2}{4\pi^2} y^{-2} +\frac{9 \zeta(7)}{16\pi^3} y^{-3} + O(e^{-4\pi |n| y})\,.
\end{align}
As mentioned already for $C_{3,1,1}$, this result was derived by studying the zero mode of the differential equation satisfied by $C_{2,1,1}(z)$ together with its large $y$ limit. We shall rederive this result from the Poincar\'e series method.

The function 
\begin{align}
\label{eq:fC211}
f(z)=C_{2,1,1}(z)- \frac{2 \pi^4}{14\, 175}E_4(z)
\end{align}
 satisfies the differential equation~\cite{DHoker:2015gmr}\footnote{Again note that the Eisenstein series in~\cite{DHoker:2015gmr} are normalised differently from those in~\eqref{eq:Eis}. }
\begin{align}
\label{eq:Lap211}
\left( \Delta - 2 \right) f(z) =  \frac{ \pi^4}{2\, 025} \left( E_4(z) - E_2(z)^2\right)\,.
\end{align}
The reason for choosing this particular combination is that the leading perturbative power $y^4$ of $y$ cancels on the right-hand side when plugging in~\eqref{eq:Eis}.

The right-hand side is problematic now because the $E_2^2$ term contains a linear term in $y$ due to~\eqref{eq:Eis} and from the seed function point of view this will introduce divergences upon Poincar\'e summation. This is very general and a special treatment must be made to discuss modular functions with source terms of the form $E_s^2$, \textit{i.e.}, squares of Eisenstein series. This was also noticed in~\cite{Klinger} where differential equations with non-linear sources $E_s E_{s'}$ were analysed using spectral theory and the case $s=s'$ had to be treated separately. This also happens for the $D^6R^4$ case discussed in Section~\ref{sec:D6R4}.

The way to proceed is to deform the differential equation~\cite{Ahlen:2018wng} and to write $E_2^2$ as the limit of $E_2 E_{2+\epsilon}$ as $\epsilon \to 0$. The deformed Laplace equation in this case can be taken as
\begin{align}
\label{eq:Lap211Def}
\left( \Delta - (2+\epsilon)(1+\epsilon) \right) f(z) =  \frac{ 2 \pi^{-\epsilon} \zeta(4+2\epsilon)}{45} \left( E_{4+\epsilon}(z) -  E_2(z) E_{2+\epsilon}(z)\right)\,.
\end{align}

Proceeding as for the $C_{3,1,1}$ case and using the Poincar\'e series~\eqref{eq:Eis} for the Eisenstein series $E_{4+\epsilon}$ and $E_{2+\epsilon}$, one can deduce that the seed function $\sigma(z)$ for $f(z)$ has to satisfy an associated differential equation that can be solved for its Fourier modes expansion $\sigma(z)=\sum_{n\in\ints} c_n(y) e^{2\pi i n x}$ by~\cite{DHoker:2015gmr,Ahlen:2018wng}
\begin{align}
\label{eq:sol211}
c_0 (y) &= \frac{\pi^{-3-\epsilon} \zeta(3)\zeta(4+2\epsilon)}{(1+\epsilon)} y^{1+\epsilon} \,,\nn\\
c_m(y) &= \frac{\pi^{-3-\epsilon}\zeta(4+2\epsilon)}{(1+\epsilon)} \sigma_{-3}(\vert m\vert) y^{1+\epsilon} e^{-2\pi |m| y}\,, \hspace{20mm} (m\neq 0)\,.
\end{align}

Once again, as anticipated in Section \ref{sec:General}, this seed function is of the general type (\ref{eq:FMgen}) discussed above when we substitute $a=-3,\, b=0,\, r=1+\epsilon$. As for the $C_{3,1,1}$ case, it is better to keep $b\neq0$ as a regulator and send it to zero only after having computed the asymptotic series (\ref{eq:IgenAsy}), finally we will send $\epsilon \to 0$.

The solution~\eqref{eq:sol211} can now be substituted into~\eqref{eq:zeroint} to obtain the zero mode of $f(z)= \sum_{n\in\ints} a_n(y) e^{2\pi i n x}$ 
and the contribution $I$ from all $c_m(y)$ with $m\neq 0$ to $a_0(y)$ is precisely captured by the general formula (\ref{eq:IgenAsy}) specialized to the present case
\begin{align}
\label{eq:I211}
I &\sim \lim_{b \to 0} \frac{\pi^{-3-\epsilon}\zeta(4+2\epsilon)}{(1+\epsilon)}   I(-3,\,b,\,1+\epsilon)\nn\\
&= \lim_{b \to 0} \frac{2^{1+2b-2\epsilon} y^{b-\epsilon}}{\pi^{2+\epsilon} \Gamma(2+\epsilon) }\Bigg[ \frac{ y}{\pi} \frac{\Gamma(b+1) \Gamma(2\epsilon-b) }{ \Gamma(\epsilon-b)}  \frac{\zeta(3+2\epsilon-3b)\zeta(4)}{\zeta(4+2\epsilon-2b)}   \nn\\
& \quad+\left(\frac{\pi}{y}\right)^{b}\sum_{n\geq0} \left(\frac{-\pi}{y}\right)^n\frac{\Gamma(n+1+2\epsilon)\zeta(-n - b)\zeta(3 - n - b) \zeta(
  1 + n +2\epsilon- b) \zeta(4 + n +2 \epsilon-b)}{n! \cdot\Gamma(n+1+\epsilon)
\zeta(4 + 2 n+2\epsilon) \zeta(4 + 2 \epsilon-2b)} \Bigg]\,,
\end{align}
  where the second term of (\ref{eq:IgenAsy}) is absent once again since it is proportional to $\zeta(a+1)$ that in the present case is $\zeta(-2)=0$.

Once more $b$ regularises the $n=2$ term and the asymptotic series terminates at $n=3$ due to the presence of the first two zetas $\zeta(-n - b)\zeta(3 - n - b)$ vanishing for $n\geq4$ in the limit $b\to 0$.

Taking the limit $b\to 0$ we produce 
\begin{align}
I \sim&\label{eq:I211Asy}\nn \frac{2^{1-2\epsilon} \zeta(3+2\epsilon)\zeta(4)}{\pi^{3+\epsilon} \Gamma(2+\epsilon) } \frac{ \Gamma(2\epsilon) }{ \Gamma(\epsilon)} y^{1-\epsilon} -\frac{4^{-\epsilon}
    \Gamma(1 + 2 \epsilon)\zeta(3) \zeta(4 + 2 \epsilon) }{
  \pi^{-2 - \epsilon} \Gamma(1 + \epsilon)\Gamma(2 + \epsilon) \zeta(2+ 2\epsilon)} \zeta(
   1 + 2 \epsilon) y^{-\epsilon}+ \\
   &+\frac{5 \zeta(5)}{12\pi} y^{-1} -\frac{\zeta(3)^2}{4\pi^2} y^{-2} +\frac{7\zeta(7)}{48 \pi^3}y^{-3}\,,
\end{align}
where we have already taken the limit $\epsilon \to 0$ in all the terms of (\ref{eq:I211}) besides the first and the $n=0$ term since they deserve some comments.

First of all, we notice in the first term of (\ref{eq:I211}) that we have the ratio $\Gamma(2\epsilon-b)/\Gamma(\epsilon-b)$ for which the two limits $b\to 0$ and $\epsilon \to 0$ do not commute!
This means that, had we started with the undeformed equation (\ref{eq:Lap211}), for which $\epsilon =0$ to begin with, we would have found the wrong coefficient. The deformation of the PDE source term $E_s^2 \to E_{s} E_{s+\epsilon}$ is crucial.

Secondly, this deformation is also crucial to regularise the divergences arising in the integral of the zero mode $c_0(y)$ of the seed function caused by the presence of a linear term in $y$ in the expansion $E_2^2$. In particular, we see that the $n=0$ term in equation (\ref{eq:I211}) produces a term in (\ref{eq:I211Asy}) proportional to  $\zeta( 1 + 2 \epsilon)$.
   
Having computed the contribution $I$ to the zero mode $a_0(y)$ in~\eqref{eq:zeroint}, we still have to consider the part coming from $c_0(y)$. With~\eqref{eq:sol211} this is given by
\begin{align}
\label{eq:c0211}
c_0(y) + y \sum_{c>0} \phi(c) \int_\reals c_0\left(\frac{y^{-1}}{c^2(1+t^2)}\right)dt &= \frac{\pi \zeta(3)}{90} y \\
&\nn\quad + \sum_{c>0} \phi(c) c^{-2-2\epsilon}  \left( \frac{4^{-\epsilon}
    \Gamma(1 + 2 \epsilon)\zeta(3) \zeta(4 + 2 \epsilon) }{
  \pi^{-2 - \epsilon} \Gamma(1 + \epsilon)\Gamma(2 + \epsilon)}  y^{-\epsilon} \right)\\
&\nn=\frac{\pi \zeta(3)}{90} y+\frac{4^{-\epsilon}
    \Gamma(1 + 2 \epsilon)\zeta(3) \zeta(4 + 2 \epsilon) }{
  \pi^{-2 - \epsilon} \Gamma(1 + \epsilon)\Gamma(2 + \epsilon)} \frac{\zeta(
   1 + 2 \epsilon) }{ \zeta(2+ 2\epsilon)}y^{-\epsilon} \,.
\end{align}
As previously stated, the deformation (\ref{eq:Lap211Def}) is essential otherwise we would have produced the divergent expression $\sum_{c>0} \phi(c) c^{-2}$ signalled by the $\zeta(1+2\epsilon)$ factor.
We note that the divergent term coming from the integral of the zero mode is matched exactly and with opposite sign by the $n=0$ term in (\ref{eq:I211Asy}).  

Combining~\eqref{eq:I211Asy} and~\eqref{eq:c0211} we cancel the divergent term and we can safely send $\epsilon \to 0$ to recover the full asymptotic zero mode perturbative expansion of $C_{2,1,1}(z)$ already presented in~\eqref{eq:a0211}, after adding back in the zero modes coming from $\frac{2 \pi^4}{14\, 175}E_4$ in order to relate $f(z)$ to $C_{2,1,1}(z)$ via~\eqref{eq:fC211}.
One can repeat a similar analysis as we did in Section \ref{sec:NP} and extract from equation (\ref{eq:I211}) using (\ref{eq:RatioZeta}) the complete non-perturbative completion of the zero mode. We have checked that this matches exactly the non-perturbative corrections recently found in \cite{DHoker:2019txf} using a completely different approach.

\section{The \texorpdfstring{$D^6R^4$}{D6R4} correction in type IIB}
\label{sec:D6R4}

We now employ the method of Section \ref{sec:General} to derive the asymptotic expansion of the $D^6R^4$ coefficient function in ten-dimensional type IIB superstring theory. Calling this function $f(z)$, it was argued in~\cite{Green:2005ba} to satisfy the differential equation
\begin{align}
\label{eq:D6R4eq}
\left( \Delta -12 \right) f(z) = - 4 \zeta(3)^2 E_{3/2}(z)^2
\end{align}
by considering compactified eleven-dimensional supergravity and making extensive use of supersymmetry.

A Poincar\'e series representation of $f(z)$ was given in~\cite{Green:2014yxa}. Here, we use a slightly different one that stems from the deformed problem studied in~\cite{Ahlen:2018wng} in order to avoid problems related to the square of the $E_{3/2}$ on the right-hand side, similar to the discussion in Section \ref{sec:211}. 

The non-zero Fourier mode of the deformed $D^6R^4$ seed is given by~\cite{Ahlen:2018wng}
\begin{align}
c_n(y) &= \frac{8\zeta(3+2\epsilon)}{1-4\epsilon^2} \sigma_{-2}(\vert n\vert) y^{1+\epsilon} \bigg((1-2\epsilon)K_2(2\pi |n| y) + \frac{5-2\epsilon}{\pi|n| y} K_3(2\pi |n| y)\nn\\
&\label{eq:defSeed}\hspace{20mm} -\frac{10-4\epsilon}{\Gamma(7/2-\epsilon)(\pi|n|y)^{1/2+\epsilon}} K_{7/2-\epsilon} (2\pi |n| y)\bigg)\,.
\end{align}
The virtue of this combination is that it is regular for $y\to 0$ and it has a convergent expansion for $y\sim 0$ of the form $y^\epsilon e^{-2\pi |n| y} \sum_{\ell\geq 0} a_\ell (4\pi |n| y)^\ell$, plus possibly $\log y$ terms. 
Note that, as derived in full detail in Appendix \ref{sec:AppBessel}, this expansion is not the usual expansion for Bessel functions at $y=0$.
It is crucial for the evaluation of the integral of the seed function, as in equation (\ref{eq:Igen}), to have both the exponential factor and a convergent expansion around the origin, the asymptotic nature of the Poincar\'e series will arise by performing the sum over the Fourier mode number $n$ \textit{after} integration and not by using the asymptotic expansion for the Bessel functions. Let us also note that we are focussing  on the expansion around $y=0$ in this discussion---even though we are ultimately interested in the asymptotic expansion of $a_0(y)$ around $y\to\infty$---is that the formula~\eqref{eq:zeroint} for $a_0(y)$ involves an `S-transformation' of $y$, that exchanges the asymptotic regimes.

Our strategy will be to write (\ref{eq:defSeed}) as a combination of the basic general building blocks~\eqref{eq:FMgen} and this can be done most conveniently by considering a shift-differential operator acting on a single term.
Using the properties of Bessel functions and confluent hypergeometric functions and after some tedious calculations, that we relegate to Appendix \ref{sec:AppBessel}, one finds that the non-zero Fourier mode (\ref{eq:defSeed}) of the deformed seed function can be written succinctly as
\begin{align}
c_n(y) = \mathcal{D} \left[\sigma_{-2}(\vert n \vert) (4\pi \vert n\vert)^\alpha y^{1+\alpha+\epsilon} e^{-2 \pi \vert n \vert y} \right]_{\alpha=0}\,,\label{eq:SeedD}
\end{align}
where $\mathcal{D}$ is a shift-differential operator in an auxiliary variable $\alpha$.
We refer to Appendix \ref{sec:AppSeed} for all the details while for the rest of the present discussion we only need to remember that $\mathcal{D}$ is function only of $\alpha$ and not of $y$ or $n$, and that its action commutes with the procedure discussed in Section \ref{sec:General} to extract the asymptotic expansion for the Poincar\'e series.

In particular we notice immediately that the form of the non-zero Fourier mode (\ref{eq:SeedD}) is precisely of the type (\ref{eq:FMgen}) considered previously. 
This means that the perturbative part to the zero mode of the $D^6R^4$ coefficient function coming from the non-zero Fourier modes of the seed function can be obtained directly from equation (\ref{eq:IgenAsy}) specialized to the present case
\begin{equation}
I \sim \mathcal{D} \left[ I(-2,\alpha,1+\alpha+\epsilon) \right]_{\alpha=0} \,.
\end{equation}

The application of the shift-differential operator in $\alpha$ after using~\eqref{eq:IgenAsy} is rather involved but straightforward, the details have been relegated to Appendix \ref{sec:AppAsymp}.
Once all the terms are collected we obtain (see equation (\ref{eq:Itotal}))
\begin{align}
\label{eq:ItotalText}
I &\sim \frac23\zeta(2)\zeta(3) y + 4\zeta(4) y^{-1} -\frac{\pi \zeta(3)^2\zeta(5)}{4\zeta(6)} y^{-2} + \frac{4\zeta(6)}{27} y^{-3}\nn\\
&\hspace{10mm} -\frac{\pi^{5/2} \Gamma(\epsilon+1/2) \zeta(1+2\epsilon)\zeta(3+2\epsilon)}{(9-6\epsilon)\Gamma(1+\epsilon)\zeta(2+2\epsilon)} y^{-\epsilon} + O(\epsilon)\,.
\end{align}
We note the occurrence of $\zeta(1+2\epsilon)$ that diverges in the limit $\epsilon\to 0$.

As before, in order to obtain the complete perturbative zero mode of the $D^6R^4$ coefficient function we also need the contributions coming from the zero-mode of the seed function, however these are easier to obtain.

The zero Fourier mode of the deformed seed is given by
\begin{align}
c_0(y) = \frac{2\zeta(3)\zeta(3+2\epsilon)}{3-6\epsilon} y^{3+\epsilon} +\frac{\pi^2 \zeta(3+2\epsilon) }{9-6\epsilon}y^{1+\epsilon}\,.
\end{align}
Its contribution to the zero mode of the Poincar\'e sum is just like for ordinary Eisenstein series. Hence we obtain
\begin{align}
\label{eqID6R400}
c_0(y) + y \sum_{c>0} \phi(c) \int_\reals c_0\left(\frac{y^{-1}}{c^2(1+t^2)}\right) dt &=  \frac{2\zeta(3)\zeta(3+2\epsilon)}{3-6\epsilon} \left[y^{3+\epsilon} +\frac{\xi(5+2\epsilon)}{\xi(6+2\epsilon)}y^{-2-\epsilon}\right] \\
&\hspace{10mm}+\frac{\pi^2 \zeta(3+2\epsilon)}{9-6\epsilon} \left[ y^{1+\epsilon} + \frac{\xi(1+2\epsilon)}{\xi(2+2\epsilon)}y^{-\epsilon}\right]\nn
\end{align}
with the completed zeta function $\xi(s) = \pi^{-s/2} \Gamma(s/2) \zeta(s)$. We see that the last term contains a $\xi(1+2\epsilon)\propto \zeta(1+2\epsilon)$ that diverges in the limit $\epsilon\to0$. As it happened in Section \ref{sec:211}, also in here this term is the reason that one has to deform the differential equation. Near $\epsilon=0$ the above expression takes the form
\begin{align}
\eqref{eqID6R400}&=\frac23 \zeta(3)^2 y^3  + \frac{\pi \zeta(3)^2\zeta(5)}{4\zeta(6)} y^{-2} + \frac23 \zeta(2)\zeta(3) y \nn\\
&\hspace{20mm}+ \frac{\pi^{5/2} \Gamma(\epsilon+1/2) \zeta(1+2\epsilon)\zeta(3+2\epsilon)}{(9-6\epsilon)\Gamma(1+\epsilon)\zeta(2+2\epsilon)} y^{-\epsilon} + O(\epsilon)\,,
\end{align}
such that the final total perturbative zero mode of the $D^6R^4$ coefficient function that arises by combining with~\eqref{eq:ItotalText} is
\begin{align}
\label{eq:ID6R4final}
\mathcal{E}_{(0,1)} &\sim \frac23 \zeta(3)^2 y^3+  \frac43\zeta(2)\zeta(3) y + 4\zeta(4) y^{-1}+ \frac{4\zeta(6)}{27} y^{-3} + O(e^{-2\pi y})
\end{align}
in agreement with~\cite{Green:2005ba,Green:2014yxa}. The three terms represent the perturbative tree-level, one-loop, two-loop and three-loop contributions to the four-graviton scattering amplitude in ten-dimensional type IIB string theory. The term proportional to $y^{-3}$ is a homogeneous solution to the differential equation~\eqref{eq:D6R4eq} satisfied by the $D^6R^4$ correction that comes out correctly of the Poincar\'e series approach. The correctness of the three-loop term was verified in a direct pure spinor calculation in~\cite{Gomez:2013sla}.

The regularisation with $\epsilon$ is important to both circumvent a divergent Poincar\'e series and to correct the $y^1$ term.  As already explained in Section \ref{sec:211}, this regularisation is necessary because the inhomogeneous Laplace equation~\eqref{eq:D6R4eq} satisfied by the $D^6R^4$ correction contains a source term, arising from the square of the $R^4$ coefficient function, that is precisely of the form $E_{3/2}^2$. From the seed function analysis this term has to be regularised via $E_{3/2} E_{3/2+\epsilon}$ as we did for the $C_{2,1,1}$ modular graph function.

\section{Conclusions}

In this paper, we have presented a method for obtaining the asymptotic expansion of certain classes of Poincar\'e series $f(z)$ whose seed Fourier modes are associated with the class~\eqref{eq:FMgen}. It is also possible to obtain non-perturbative information from this asymptotic expansion using resurgent analysis as we have demonstrated in several examples.

At the moment the focus of our studies has been entirely devoted to the derivation of the zero mode sector. An obvious future direction is to extend our analysis to the non-zero modes and derive their perturbative and non-perturbative expansions starting from the integral form (\ref{eq:NZmodes}).

Another task yet to be completed is the reconstruction of the non-perturbative corrections to the $D^6R^4$ coefficient function by combining the use of the shift operator introduced to rewrite the Fourier modes (\ref{eq:SeedD}) together with a Cheshire-cat type of resurgence similar to the discussion in Section \ref{sec:NP} for the simpler setup of the $C_{3,1,1}$ modular graph function.

Furthermore, even within the class of modular graph functions discussed in the present paper, it is conceivable that from our asymptotic expansion (\ref{eq:IgenAsy}) it will be possible to write the general expression of the non-perturbative corrections, reconstructing them entirely out of the perturbative data. For the two cases $C_{3,1,1}$ and $C_{2,1,1}$ discussed in here we have checked that these non-perturbative corrections match exactly the one derived recently in \cite{DHoker:2019txf} using a completely different approach.

We have little doubt that our method can be applied to many more modular graph functions, for example those in the class $C_{a,b,c}$ studied in~\cite{DHoker:2015gmr} or the tetrahedral functions studied in~\cite{Kleinschmidt:2017ege} and whose Laplace equations are known. Since the Laplace equation for a given  value of $(a,b,c)$ higher than the examples considered here can generally involve also sources involving other $C_{a',b',c'}$ functions (before diagonalisation of the Laplacian), we expect the corresponding Fourier modes to be more involved but still derivable from~\eqref{eq:FMgen}, presumably up to the action of a differential operator. 
It is also extremely interesting to understand how to extend our analysis to different type of multiplicative functions, besides the divisor sum $\sigma_{a}$, appearing in the seed function (\ref{eq:FMgen}) and whether or not such cases exist within string theory.

There are several possible generalisations of the present analysis presented here. Besides an extension to the analysis of the non-zero modes $f(z)$, one tantalising avenue seems to be to use the methods to investigate higher-derivative terms in the type IIB effective action starting from $D^8R^4$ whose exact form is currently unknown. Laplace equations for $D^8R^4$ and, in particular, $D^{10}R^4$ have been proposed in the literature~\cite{Green:2005ba,Basu:2006cs} and it would be interesting to analyse them via the Poincar\'e series approach and study their perturbative terms.

It is also conceivable to extend the present strategy beyond $SL(2,\ints)$ to higher rank groups. According to the two instances in the introduction, there are two different generalisations: either going to higher genus world-sheets (which means $Sp(2g,\ints)$ for low genus $g$) or going to higher rank U-duality groups (which means $E_n(\ints)$ for compactification of type II on a space-time torus $T^{n-1}$). Modular graph functions for world-sheets of genus two and higher have been explored recently~\cite{DHoker:2013fcx,DHoker:2014oxd,DHoker:2017pvk,Basu:2018eep,DHoker:2018mys,Basu:2018bde} with a focus on obtaining the perturbative terms in the non-separating divisor limit of the Riemann surface that is similar to the asymptotic expansion studied in the present paper. 

For higher rank U-duality groups $E_n(\ints)$, there is a fair amount known for the lowest derivative corrections $R^4$ and $D^4R^4$, see~\cite{Green:2010kv,Green:2011vz,Pioline:2010kb,Fleig:2012xa}, where the solutions can be given in terms of parabolic Eisenstein series, \textit{i.e.}, Poincar\'e sums of characters of a parabolic subgroup. As Eisenstein series satisfy homogeneous differential equations and since there are many methods available for analysing them, these cases have been treated in detail. The situation is less clear starting from $D^6R^4$ where the differential equation implied by supersymmetry again is inhomogeneous~\cite{Green:2005ba,Pioline:2015yea,Bossard:2015uga}. For some higher rank cases, solutions have been proposed in the literature based on different methods~\cite{Pioline:2015yea,Bossard:2015foa,Ahlen:2018wng}, however, their equivalence and their full physical content in terms of an asymptotic expansion are in general not known. We hope that the methods of the present paper can also help to study these functions.

\subsection*{Acknowledgements}

We would like to thank Jens Funke, Herbert Gangl, Jan Gerken and Oliver Schlotterer for useful discussions. DD would like to thank the Albert Einstein Institute and in particular Hermann Nicolai for the hospitality and support during the various stages of this project. AK gratefully acknowledges support from the Simons Center for Geometry and Physics, Stony Brook University at which part of the research for this paper was performed.

\appendix

\section{Useful identities}

\subsection{From Poincar\'e series to Kloosterman sums}
\label{app:Kloost}

In this appendix, we briefly review how to obtain the relation~\eqref{eq:NZmodes} between the Fourier coefficients of the seed of a Poincar\'e and those of the summed function. We follow~\cite{Iwaniec,Fleig:2015vky} to which we direct the reader for additional details and references. 

The Poincar\'e sum~\eqref{eq:PSf} is a sum over cosets in $B(\ints)\backslash SL(2,\ints)$. These can be represented by matrices
\begin{align}
\begin{pmatrix} a & b \\ c&  d\end{pmatrix}
\end{align}
with $c\in\ints$ and $d>0$ \textit{coprime}, where choosing a representative of the coset means choosing any fixed solution for $a$ and $b$ of the condition $ad-bc=1$. The sum over coset classes with coprime $c$ and $d$ can be further refined by grouping the terms in $d$ modulo $c$ using a further right quotient $B(\ints)\backslash SL(2,\ints)/B(\ints)$ as follows. Multiplying by elements from $B(\ints)$ on the right means
\begin{align}
\begin{pmatrix} a & b \\ c&  d\end{pmatrix}\begin{pmatrix} 1 & k \\ 0&  1\end{pmatrix}
=\begin{pmatrix} a & b+ ak \\ c&  d+ck \end{pmatrix}\,,
\end{align}
so one may sum over all $k\in\ints$ and all elements in $\ints/c\ints$ that coprime with $c$ as long as $c\neq0$. This is the set $(\ints/c\ints)^\times$ that appears in many places in this article. For $c=0$, there is only the term $d=1$ and the coset is represented by the identity element. Therefore the Poincar\'e sum~\eqref{eq:PSf} is
\begin{align}
f(z) = \sigma(z) + \sum_{c>0} \sum_{d\in (\ints/c\ints)^\times} \sum_{k\in\ints} \sigma\left( \frac{a}{c} - \frac{1}{c(c(z+k)+d)} \right)\,,
\end{align}
where we have rewritten the argument of $\sigma$ using $ad-bc=1$ with $a$ any fixed solution. The sum over $c\neq 0$ is restricted to positive numbers as the right quotient by $B(\ints)$ also includes a possible sign change. 

Performing a Poisson resummation on $k\in\ints$ leads to
\begin{align}
f(z) = \sigma(z) + \sum_{c>0} \sum_{d\in (\ints/c\ints)^\times} \sum_{n\in\ints} \int_{\reals} e^{-2\pi i n \omega} \sigma\left( \frac{a}{c} - \frac{1}{c(c(z+\omega)+d)} \right) d\omega\,.
\end{align}
Shifting the integration variable $\omega\to \omega+x +d/c$ then yields with $z=x+iy$
\begin{align}
f(z) &= \sigma(z) + \sum_{n\in\ints} e^{2\pi i n x} \sum_{c>0} \sum_{d\in (\ints/c\ints)^\times} e^{2\pi i n d/c} \int_{\reals} e^{-2\pi i n \omega} \sigma\left( \frac{a}{c} - \frac{1}{c^2(\omega+iy)} \right) d\omega \nn\\
&=\sigma(z) \\
&\quad+ \sum_{n\in\ints} e^{2\pi i n x} \sum_{c>0} \sum_{d\in (\ints/c\ints)^\times} e^{2\pi i n d/c} \int_{\reals} e^{-2\pi i n \omega} \sum_{m\in\ints} e^{2\pi i m a/c} e^{-2\pi i m \frac{\omega}{c^2(\omega^2+y^2)}} c_m\left(\frac{y}{c^2 (\omega^2+y^2)}\right)d\omega \,,\nn
\end{align}
where we have inserted the Fourier expansion $\sigma(z) = \sum_{m\in\ints} c_m(y)e^{2\pi i m x} $ at the given argument.

{}From the above equation we can read off the Fourier modes of $f(z)=\sum_{n\in\ints} a_n(y) e^{2\pi in x}$ as
\begin{align}
a_n(y) = \sigma_n(y) + \sum_{c>0} \sum_{d\in(\ints/c\ints)^\times} \sum_{m\in\ints} e^{2\pi i n d/c + 2\pi i m a/c} \int_\reals e^{-2\pi i n \omega -2\pi i m \frac{\omega}{c^2(y^2+\omega^2)}}c_m\left(\frac{y}{c^2(y^2+\omega)}\right) d \omega\,.
\end{align}
Replacing $d$ by $q$ one arrives at~\eqref{eq:NZmodes}.

\subsection{Integrals of Fourier modes}
\label{app:intFM}

The Fourier modes we encounter are combinations of terms of the form $c_n(y)\propto y^r e^{-2\pi n y} $ for some power $r$. For the zero modes in~\eqref{eq:zeroint} we then have to evaluate integrals of the type
\begin{align}
\int_\reals e^{-2\pi  m \frac{1+i t}{y c^2 (1+t^2)}} \frac{1}{(1+t^2)^r} dt\,.
\end{align}
Expanding the exponential, the individual terms are in the class (for $\Re(a+b)>1$)
\begin{align}
\int_\reals \frac1{(1+it)^a} \frac1{(1-it)^b} dt = 2^{2-a-b} \pi \frac{\Gamma(a+b-1)}{\Gamma(a)\Gamma(b)}\,,
\end{align}
such that
\begin{align}
\label{eq:auxint1}
\int_\reals e^{-2\pi m \frac{1+i t}{y c^2 (1+t^2)}} \frac{1}{(1+t^2)^r} dt 
= \frac{\pi}{4^{r-1}\Gamma(r)} \sum_{k\geq 0} \frac{(-\pi m y^{-1} c^{-2})^k}{k!} \frac{\Gamma(2r+k-1)}{\Gamma(k+r)}\,,
\end{align}
valid for $\Re(r)>1/2$.
For integer $r$ this is a polynomial of degree $r-1$ in $\frac{\pi m}{y c^2}$ times $\exp(-\frac{\pi m}{y c^2})$. For generic $r$ we can rewrite the result using the shift operator $D_\alpha = e^{\partial_\alpha}$ that satisfies $D_\alpha^k f(\alpha) = f(\alpha+k)$. Then the formula becomes
\begin{align}
\int_\reals e^{-2\pi  m \frac{1+i t}{y c^2 (1+t^2)}} \frac{1}{(1+t^2)^r} dt 
= \left.\frac{\pi}{4^{r-1}\Gamma(r)} \exp\left( - \frac{\pi m}{yc^2} D_\alpha\right) \frac{\Gamma(2r+\alpha-1)}{\Gamma(\alpha+r)}\right|_{\alpha=0}\,.
\end{align}

Another integral that will be useful is
\begin{align}
\label{eq:auxint2}
\int_0^\infty \sum_{n>0} \theta^{nhq} n^{-s} t^b \sum_{k \geq 0} \frac{(-nt)^k}{k!} \frac{\Gamma(2r+k-1)}{\Gamma(r+k)} dt
= \frac{\Gamma(b+1) \Gamma(2r-b-2)}{\Gamma(r-b-1)} \Li_{s+b+1} (\theta^{hq}) 
\end{align}
that can be computed using the shift operator $D_\alpha$ and also uses
\begin{align}
\int_0^\infty t^b \Li_s(\theta^{hq} e^{-t}) dt = \Gamma(b+1) \Li_{s+b+1}(\theta^{hq})\,.
\end{align}

\subsection{Sums of polylogarithms}
\label{app:PLsums}

In this appendix, we consider sums of the form
\begin{align}
\label{eq:zetaPL}
 \sum_{c>0} c^{-s} \sum_{q\in(\ints/c\ints)^\times} \sum_{h=1}^{c} \zeta\big(1-k,\frac{h}{c}\big) \Li_{n}(\theta^{hq})\,,
\end{align}
where $\theta= e^{2\pi i /c}$ is a primitive $c$-th root of unity.  The parameters $k$ and $n$ need not be integers in this expression and in fact many expressions become singular when they are. 

First, we note that the Hurwitz zeta function can be rewritten according to~\cite{DLMF}
\begin{align}
\zeta\big(1-k,\frac{h}{c}\big) = \frac{\Gamma(k)}{(2\pi)^k} \left( i^{-k} \Li_k (\theta^h) + i^k \Li_k(\theta^{-h})\right)\,.
\end{align}
The two terms are related by complex conjugation, so it is sufficient to consider one of them. Therefore the basic object we are facing is
\begin{align}
 \sum_{c>0} c^{-s}  \sum_{q\in(\ints/c\ints)^\times} \sum_{h=0}^{c-1} \Li_k(\theta^h) \Li_n(\theta^{hq})\,,
\end{align}
where we have shifted the $h$-summation using the periodicity of $\theta^h$.

Under the (numerically verified) assumption that the object obtained by summing over $q$ and $h$ and dividing by $\Li_k(1)\Li_n(1)$ is multiplicative in $c$, it is sufficient to consider the case $c=p^\ell$, where $p$ is a prime and $\ell$ an integer. In that case one has to determine
\begin{align}
\mathcal{S}_{k,n}(p^\ell) 
&=\sum_{h=0}^{p^\ell-1} \sum_{q\in(\ints/p^\ell\ints)^\times}  \Li_k(\theta^h) \Li_n(\theta^{hq})\\
&= \sum_{m=0}^{\ell-1} \sum_{a\in (\ints/p^{\ell-m}\ints)^\times} \sum_{q \in (\ints/p^\ell \ints)^\times} \Li_k (\theta^{ap^m}) \Li_n (\theta^{a p^m q}) + \Li_k(1) \Li_n(1) \phi (p^\ell)\,,\nn
\end{align}
where we have grouped the sum over $h$ into different classes. Now one uses that for $\gcd ( a, p^{\ell-m}) =1$ the second factor in the sum is independent of $a$ and yields
\begin{align}
\sum_{q\in (\ints/p^\ell \ints)^\times} \Li_n(\theta^{a p^m q} ) &= \sum_{q\in (\ints/p^\ell \ints)^\times} \Li_n(\theta^{ p^m q} )= \Li_n(1) p^m \left( p^{(\ell-m) (1-n)} - p^{(\ell-m-1)(1-n)}\right)\,,
\end{align}
while the first factor leads to
\begin{align}
\sum_{a\in (\ints/p^{\ell-m}\ints)^\times} \Li_k(\theta^{ap^m}) = \Li_k(1) \left( p^{(\ell-m)(1-k)} - p^{(\ell-m-1)(1-k)}\right)\,.
\end{align}
We obtain therefore
\begin{align}
&\hspace{10mm} \frac1{\Li_k(1)\Li_n(1)}\sum_{h=0}^{p^\ell-1} \sum_{q\in(\ints/p^\ell\ints)^\times}  \Li_k(\theta^h) \Li_n(\theta^{hq})\nn\\
&= \sum_{m=0}^{\ell-1}\left[p^m \left( p^{(\ell-m) (1-n)} - p^{(\ell-m-1)(1-n)}\right)\left( p^{(\ell-m)(1-k)} - p^{(\ell-m-1)(1-k)}\right)\right] +\phi(p^{\ell}) \nn\\
&= (1-p^{k-1}) (1-p^{n-1})  p^{-\ell(k+n-2)} \frac{1-p^{\ell(k+n-1)}}{1-p^{k+n-1}}
 + \phi(p^{\ell}) \,.
\end{align}

For the Dirichlet series written as an Euler product we next need to form 
\begin{align}
\sum_{\ell\geq 0} \mathcal{S}_{k,n}(p^\ell) p^{-\ell s} &= \Li_k(1) \Li_n(1) \frac{(1-p^{1-k-s})(1-p^{1-n-s})}{(1-p^{2-n-k-s})(1-p^{1-s})}\,,
\end{align}
leading for sufficiently large $\Re(s)$ to 
\begin{align}
 \sum_{c>0} c^{-s}  \sum_{q\in(\ints/c\ints)^\times} \sum_{h=0}^{c-1} \Li_k(\theta^h) \Li_n(\theta^{hq})
 = \zeta(k) \zeta(n)\frac{\zeta(n+k+s-2)\zeta(s-1)}{\zeta(k+s-1)\zeta(n+s-1)} \,.
\end{align}
where we have also substituted $\Li_k(1) = \zeta(k)$ to continue $\Li_k(1)$ to most values of $k$. As said before this calculation requires $k$ and $n$ to be analytically continued and the above formula works well to determine limiting values.

Returning to the original expression~\eqref{eq:zetaPL} we then find 
\begin{align}
\label{eq:zetaPL2}
 \sum_{c>0} c^{-s} \!\!\!\! \sum_{q\in(\ints/c\ints)^\times} \sum_{h=1}^{c} \zeta\big(1-k,\frac{h}{c}\big) \Li_{n}(\theta^{hq}) &= \frac{\Gamma(k)}{(2\pi i)^k}  \zeta(k) \zeta(n)\frac{\zeta(n+k+s-2)\zeta(s-1)}{\zeta(k+s-1)\zeta(n+s-1)}  + \text{c.c.}\nn\\
 &= \frac{\zeta(1-k) \zeta(n)\zeta(n+k+s-2)\zeta(s-1)}{\zeta(k+s-1)\zeta(n+s-1)}  \,.
\end{align}
Here, $\text{c.c.}$ stands for the complex conjugate and we have used the functional equations for $\zeta(k)$ and $\Gamma(k)$ for the simplified expression.

We also note
\begin{align}
\label{eq:PL1}
\sum_{c>0} c^{-s} \sum_{q\in(\ints/c\ints)^\times} \sum_{h=1}^{c} \Li_n(\theta^{hq}) = \sum_{c>0} c^{-s+1-n} \phi(c) \Li_n(1) = \frac{\zeta(n) \zeta(s+n-2)}{\zeta(s+n-1)}\,.
\end{align}

When taking limits it can be useful to remember that the derivative of the Riemann zeta function at negative even integers satisfies 
\begin{align}
\zeta'(-2n) = (-1)^n \frac{(2n)!}{2 (2\pi)^{2n}} \zeta(2n+1)\,.
\end{align}

\section{Expanding the \texorpdfstring{$D^6R^4$}{D6R4} seed modes}

In this appendix, we perform the rewriting of the Fourier mode of $D^6R^4$ seed function presented in~\eqref{eq:defSeed} and derive its contribution to the asymptotic expansion of the zero mode of the $D^6R^4$ coefficient itself.

\subsection{Bessel functions and confluent hypergeometric functions}
\label{sec:AppBessel}

The modified Bessel function can be written as
\begin{align}
\label{eq:KtoU}
K_s(z) = \sqrt{\pi} e^{-z} (2z)^s  U(s+\tfrac12,2s+1; 2z)
\end{align}
in terms of the confluent hypergeometric function $U(a,b;z)$ that is a variant of Kummer's function
\begin{align}
\label{eq:KummerM}
M(a,b;z) = {}_1F_1(a,b;z) = \sum_{\ell\geq 0} \frac{(a)_\ell}{(b)_\ell} \frac{z^\ell}{\ell!} \,,
\end{align}
defined in terms of the \textit{rising} Pochhammer symbols $(a)_\ell = a \cdot (a+1) \cdots (a+\ell-1) = \frac{\Gamma(a+\ell)}{\Gamma(a)}$. 
 For $b\notin \ints$ one has the relation~\cite{DLMF}
\begin{align}
\label{eq:UtoM}
U(a,b;z) = \frac{\Gamma(1-b)}{\Gamma(a-b+1)} M(a,b;z) + z^{1-b} \frac{\Gamma(b-1)}{\Gamma(a)} M(a-b+1,2-b;z)\,.
\end{align}
Putting this back into~\eqref{eq:KtoU} makes the symmetry $K_s(z) = K_{-s}(z)$ manifest. The second term in~\eqref{eq:UtoM} contains a finite number of singular terms (in $z$) for $b>1$ but $b\notin \ints$.

We also require the expansion of the confluent hypergeometric function $U(a,n+1;z)$ around $z= 0$ when $n\in \ints_{\geq 0}$. This is given by~\cite{DLMF}
\begin{align}
\label{eq:Uint}
U(a,n+1;z) = \tilde{U}(a,n+1;z) + \frac{\Gamma(n)}{\Gamma(a)} z^{-n} \sum_{\ell=0}^{n-1} \frac{(a-n)_\ell}{(1-n)_\ell} \frac{z^\ell}{\ell!}
\end{align}
where 
\begin{align}
\label{eq:Utilde}
&\hspace{10mm}\tilde{U}(a,n+1;z) \\
&= \frac{(-1)^{n+1}}{\Gamma(a)\Gamma(a-n)} \sum_{\ell\geq 0} \frac{ \Gamma(a+\ell) z^\ell}{\Gamma(n+1+\ell)\Gamma(1+\ell)} \left( \log(z) + \psi(a+\ell) -\psi(n+\ell+1)-\psi(\ell+1)\right)\nn
\end{align}
contains the regular and log terms and the digamma function is the logarithmic derivative of the gamma function according to $\psi(x) = \Gamma'(x) /\Gamma(x)$.
Using formulas~\eqref{eq:KtoU},~\eqref{eq:UtoM} and~\eqref{eq:Uint} we can obtain the expansion of $K_s(z)$ of the form
\begin{align}
K_s(z) = e^{-z} \times \textrm{(shifted Laurent series around $z=0$)}\,.
\end{align}
The lowest power that occurs in the series is $z^{-s}$ for any real $s>0$.

We note that we can rewrite~\eqref{eq:Utilde} by introducing an auxiliary parameter $\alpha$ as in appendix~\ref{app:intFM}. In this way one obtains with the shift operator $D_\alpha= e^{\partial_\alpha}$ the following compact expression for the regular part $\tilde{U}(a,n+1;z)$:
\begin{align}
\label{eq:tUdiff}
\tilde{U}(a,n+1;z) &=  \frac{(-1)^{n+1}}{\Gamma(a)\Gamma(a-n)} \sum_{\ell\geq 0} \partial_\alpha \left[ \frac{\Gamma(a+\ell+\alpha) z^{\ell+\alpha}}{\Gamma(n+\ell+1+\alpha)\Gamma(\ell+1+\alpha)}\right]_{\alpha=0}\nn\\
&=  \frac{(-1)^{n+1}}{\Gamma(a)\Gamma(a-n)} \sum_{\ell\geq 0} D_\alpha^\ell \partial_\alpha \left[ \frac{\Gamma(a+\alpha) z^{\alpha}}{\Gamma(n+1+\alpha)\Gamma(1+\alpha)}\right]_{\alpha=0}\nn\\
&=  \frac{(-1)^{n+1}}{\Gamma(a)\Gamma(a-n)} \frac{1}{1-D_\alpha} \partial_\alpha \left[ \frac{\Gamma(a+\alpha) z^{\alpha}}{\Gamma(n+1+\alpha)\Gamma(1+\alpha)}\right]_{\alpha=0}\,.
\end{align}

\subsection{Rewriting the \texorpdfstring{$D^6R^4$}{D6R4} seed}
\label{sec:AppSeed}
The specific combination appearing in the $D^6R^4$ seed Fourier mode is 
\begin{align}
\label{eq:cnD6R4}
c_n(y) &= \frac{8\zeta(3+2\epsilon)}{1-4\epsilon^2} \sigma_{-2}(n)(2\pi|n|)^{-1-\epsilon} \\
&\hspace{10mm} \times z^{1+\epsilon} \left[ (1-2\epsilon) K_2(z) + \frac{10-4\epsilon}{z} K_3(z) - \frac{(10-4\epsilon) 2^{1/2+\epsilon}}{\Gamma(7/2-\epsilon)z^{1/2+\epsilon}} K_{7/2-\epsilon}(z)\right]\,,\nn
\end{align}
where we use $z=2\pi|n| y$. 

We shall first rewrite the second line in the form $e^{-z} \times \textrm{(convergent power series in $z$)}$, showing along the way that there are no negative powers of $z$ appearing in this particular combination.
As a second step we shall write the whole expression as the action of a differential operator acting on a simpler term.

The expansions of the various Bessel functions near $z=0$ are
\begin{align}
K_2(z) &= \sqrt{\pi} e^{-z}  \left[ (2z)^2 \tilde{U}(\tfrac{5}{2},5;2z) + \frac{\Gamma(4)}{\Gamma(5/2)} (2z)^{-2}\left(1+z +\frac{z^2}{4} - \frac{z^3}{12}  \right)\right]\,,\nn\\
K_3(z) &= \sqrt{\pi} e^{-z}  \left[(2z)^3\tilde{U}(\tfrac{7}{2},7;2z) + \frac{\Gamma(6)}{\Gamma(7/2)} (2z)^{-3}\left(1+z  +\frac{3z^2}{8} + \frac{z^3}{24}  -\frac{z^4}{192}+\frac{z^5}{320}\right)\right]\,,\nn\\
K_{7/2-\epsilon}(z) &= \sqrt{\pi} e^{-z} \bigg[(2z)^{7/2-\epsilon}  \frac{\Gamma(2\epsilon-7)}{\Gamma(\epsilon-3)}M(4-\epsilon,8-2\epsilon;2z) \\
&\hspace{50mm}+ (2z)^{\epsilon-7/2} \frac{\Gamma(7-2\epsilon)}{\Gamma(4-\epsilon)} M(\epsilon-3,2\epsilon-6;2z) \bigg]\nn\\
&= \sqrt{\pi} e^{-z} \Bigg[(2z)^{7/2-\epsilon}  \frac{\Gamma(2\epsilon-7)}{\Gamma(\epsilon-3)}M(4-\epsilon,8-2\epsilon;2z) \nn\\
&\hspace{5mm}+ (2z)^{\epsilon-7/2} \frac{\Gamma(7-2\epsilon)}{\Gamma(4-\epsilon)} \left( 1 + z + \frac{2-\epsilon}{5-2\epsilon}z^2 + \frac{1-\epsilon}{15-6\epsilon} z^3+\sum_{\ell> 3} \frac{(\epsilon-3)_\ell}{(2\epsilon-6)_\ell} \frac{(2z)^\ell}{\ell!}  \right) \Bigg]\,.\nn
\end{align}
Note that, as stressed above, this is not the usual asymptotic expansion of the Bessel functions around $z=0$ where we have stripped away the exponential factor, necessary for the convergence of the integral (\ref{eq:zeroint}).

Taking into account also the pre-factors of the second line of~\eqref{eq:cnD6R4}, the possible singular and constant terms are of the orders $z^{-3+\epsilon}$, $z^{-2+\epsilon}$, $z^{-1+\epsilon}$ and $z^\epsilon$ with coefficients (after dropping the common $\sqrt{\pi}e^{-z}$) :
\begin{align}
z^{-3+\epsilon}&:\hspace{5mm} (10-4\epsilon)  \frac{\Gamma(6)}{\Gamma(7/2)} 2^{-3}  - \frac{10-4\epsilon}{\Gamma(7/2-\epsilon)} 2^{2\epsilon -3} \frac{\Gamma(7-2\epsilon)}{\Gamma(4-\epsilon)} =0\,,\nn\\
z^{-2+\epsilon}&:\hspace{5mm} (10-4\epsilon)  \frac{\Gamma(6)}{\Gamma(7/2)} 2^{-3}  - \frac{10-4\epsilon}{\Gamma(7/2-\epsilon)} 2^{2\epsilon -3} \frac{\Gamma(7-2\epsilon)}{\Gamma(4-\epsilon)} =0\,,\\
z^{-1+\epsilon}&:\hspace{5mm} (1-2\epsilon) \frac{\Gamma(4)}{\Gamma(5/2)} 2^{-2} + \frac38(10-4\epsilon)  \frac{\Gamma(6)}{\Gamma(7/2)} 2^{-3}  - \frac{2-\epsilon}{5-2\epsilon}\frac{10-4\epsilon}{\Gamma(7/2-\epsilon)} 2^{2\epsilon -3} \frac{\Gamma(7-2\epsilon)}{\Gamma(4-\epsilon)} =0\,,\nn\\
z^{\epsilon}&:\hspace{5mm} (1-2\epsilon) \frac{\Gamma(4)}{\Gamma(5/2)} 2^{-2} + \frac1{24}(10-4\epsilon)  \frac{\Gamma(6)}{\Gamma(7/2)} 2^{-3}  - \frac{1-\epsilon}{15-6\epsilon}\frac{10-4\epsilon}{\Gamma(7/2-\epsilon)} 2^{2\epsilon -3} \frac{\Gamma(7-2\epsilon)}{\Gamma(4-\epsilon)} =0\,.\nn
\end{align}
Therefore, the non-zero Fourier mode~\eqref{eq:cnD6R4} of the seed of the $D^6R^4$ function can be written in an expansion where every term is at least of the order $z^{1+\epsilon}$:
\begin{align}
\label{eq:cnD6R41}
c_n(y) &= \frac{8\sqrt{\pi}\zeta(3+2\epsilon)}{1-4\epsilon^2} \sigma_{-2}(n)(2\pi|n|)^{-1-\epsilon}  e^{-z}\nn\\
&\quad\times\bigg[ 4(1-2\epsilon) z^{3+\epsilon} \tilde{U}(\tfrac{5}{2},5;2z) + \frac{\Gamma(4)}{4\Gamma(5/2)}(1-2\epsilon) \left( \frac{z^{1+\epsilon}}{4}-\frac{z^{2+\epsilon}}{12} \right)\nn\\
&\hspace{10mm} + 8  (10-4\epsilon) z^{3+\epsilon} \tilde{U}(\tfrac72,7;2z) + \frac{\Gamma(6)}{8\Gamma(7/2)}(10-4\epsilon) \left( - \frac{z^{1+\epsilon}}{192} + \frac{z^{2+\epsilon}}{320} \right)\nn\\
&\hspace{10mm} - 16\frac{(10-4\epsilon)\Gamma(2\epsilon-7)}{\Gamma(7/2-\epsilon)\Gamma(\epsilon-3)} z^{4-\epsilon} M(4-\epsilon,8-2\epsilon;2z) -8\frac{10-4\epsilon}{\sqrt{\pi}} z^{-3+\epsilon} \sum_{\ell>3} \frac{(\epsilon-3)_\ell}{(2\epsilon-6)_\ell} \frac{(2z)^\ell}{\ell!}\Bigg]\,,\nn\\
&= \frac{8\sqrt{\pi}\zeta(3+2\epsilon)}{1-4\epsilon^2} \sigma_{-2}(n)(4\pi|n|)^{-1-\epsilon}  e^{-z}\\
&\quad\times\!\Bigg[ (1-2\epsilon) (2z)^{3+\epsilon} \tilde{U}(\tfrac{5}{2},5;2z)\! + \!  (20-8\epsilon) (2z)^{3+\epsilon} \tilde{U}(\tfrac72,7;2z) \!+\! \frac{1-10\epsilon}{12\sqrt{\pi}} (2z)^{1+\epsilon} \!+\! \frac{5+14\epsilon}{120\sqrt{\pi}}(2z)^{2+\epsilon}\nn\\
&\hspace{10mm} +2^{2+2\epsilon}\frac{\Gamma(\epsilon-3/2)}{\pi} \bigg( \frac{\Gamma(1+\epsilon)}{\Gamma(2\epsilon-2)\Gamma(5)} (2z)^{1+\epsilon} + \frac{\Gamma(2+\epsilon)}{\Gamma(2\epsilon-1)\Gamma(6)} (2z)^{2+\epsilon} + \frac{\Gamma(3+\epsilon)}{\Gamma(2\epsilon)\Gamma(7)} (2z)^{3+\epsilon}\bigg)\nn\\
&\hspace{10mm} - 2^{2+2\epsilon} \frac{\Gamma(\epsilon-3/2)}{\pi} \sum_{\ell\geq 0} \left(   \frac{\Gamma(4+\ell-\epsilon)(2z)^{4+\ell-\epsilon}}{\Gamma(8+\ell-2\epsilon)\Gamma(1+\ell)}  - \frac{\Gamma(4+\ell+\epsilon)(2z)^{4+\ell+\epsilon} }{\Gamma(8+\ell)\Gamma(1+\ell+2\epsilon)} \right) \Bigg]\nn
\end{align} 
where we have also inserted the expansion of Kummer's function~\eqref{eq:KummerM}. Note that the above expression has a smooth limit $\epsilon\to 0$ in which the sum over $\ell$ disappears, reflecting the fact that $K_{7/2-\epsilon}\to K_{7/2}$ with a finite expansion, and the three terms in the second line also disappear such that
\begin{align}
c_n(y) &\underset{\epsilon\to 0}{\longrightarrow} 8\sqrt{\pi}\zeta(3) \sigma_{-2}(n)(4\pi|n|)^{-1}  e^{-z}
\bigg[  (2z)^{3} \tilde{U}(\tfrac{5}{2},5;2z) + 20 (2z)^{3} \tilde{U}(\tfrac72,7;2z) + \frac{2z}{12\sqrt{\pi}} +\frac{(2z)^2}{24\sqrt{\pi}}\bigg]\,.
\end{align}

We further simplify~\eqref{eq:cnD6R41} by writing it as the application of a differential on a simpler function. This was already done for the $\tilde{U}$ functions in~\eqref{eq:tUdiff}. 

We start with the terms involving $\tilde{U}$ and rewrite the relevant part of~\eqref{eq:cnD6R41} as
\begin{align}
&\quad(1-2\epsilon) (2z)^{3+\epsilon} \tilde{U}(\tfrac52,5;2z) + (20-8\epsilon) (2z)^{3+\epsilon} \tilde{U}(\tfrac72,7;2z)\\
&= -\frac{1}{1-D_\alpha} \partial_\alpha\left[\frac{(1-2\epsilon) (2z)^{3+\epsilon}}{\Gamma(5/2)\Gamma(-3/2)} \frac{\Gamma(5/2+\alpha) (2z)^\alpha}{\Gamma(5+\alpha)\Gamma(1+\alpha)} + \frac{(20-8\epsilon) (2z)^{3+\epsilon}}{\Gamma(7/2)\Gamma(-5/2)} \frac{\Gamma(7/2+\alpha)(2z)^\alpha}{\Gamma(7+\alpha)\Gamma(1+\alpha)} \right]_{\alpha=0}\nn\\
&= - \frac{1}{\pi}\frac{D_\alpha^2}{1-D_\alpha}\partial_\alpha\left[ \left(\alpha^2-13\alpha+2 -2 \epsilon (\alpha^2+3\alpha+10) \right) \frac{\Gamma(1/2+\alpha)}{\Gamma(5+\alpha)\Gamma(\alpha-1)}(2z)^{1+\alpha+\epsilon}\right]_{\alpha=0}\,.\nn
\end{align}
We note also that
\begin{align}
&\quad\quad -\frac{1}{\pi} (1+D_\alpha)\partial_\alpha\left[ \left(\alpha^2-13\alpha+2 -2 \epsilon (\alpha^2+3\alpha+10) \right) \frac{\Gamma(1/2+\alpha)}{\Gamma(5+\alpha)\Gamma(\alpha-1)}(2z)^{1+\alpha+\epsilon}\right]_{\alpha=0} \nn\\
&= \frac{1-10\epsilon}{12\sqrt{\pi}} (2z)^{1+\epsilon} + \frac{5+14\epsilon}{120\sqrt{\pi}}(2z)^{2+\epsilon}\,,
\end{align}
so that the first line of the square brackets in~\eqref{eq:cnD6R41} can be rewritten as
\begin{align}
&\quad(1-2\epsilon) (2z)^{3+\epsilon} \tilde{U}(\tfrac52,5;2z) + (20-8\epsilon) (2z)^{3+\epsilon} \tilde{U}(\tfrac72,7;2z)+\frac{1-10\epsilon}{12\sqrt{\pi}} (2z)^{1+\epsilon} + \frac{5+14\epsilon}{120\sqrt{\pi}}(2z)^{2+\epsilon}
\nn\\
&= -\frac{1}{\pi}\frac1{1-D_\alpha}\partial_\alpha \left[\left(\alpha^2-13\alpha+2 -2 \epsilon (\alpha^2+3\alpha+10) \right) \frac{\Gamma(1/2+\alpha)}{\Gamma(5+\alpha)\Gamma(\alpha-1)}z^{1+\alpha+\epsilon}\right]_{\alpha=0}\,.
\end{align}

The three contributions in the second line of~\eqref{eq:cnD6R41} can be written as
\begin{align}
&\hspace{5mm} 2^{2+2\epsilon}\frac{\Gamma(\epsilon-3/2)}{\pi} \bigg( \frac{\Gamma(1+\epsilon)}{\Gamma(2\epsilon-2)\Gamma(5)} (2z)^{1+\epsilon} + \frac{\Gamma(2+\epsilon)}{\Gamma(2\epsilon-1)\Gamma(6)} (2z)^{2+\epsilon} + \frac{\Gamma(3+\epsilon)}{\Gamma(2\epsilon)\Gamma(7)} (2z)^{3+\epsilon}\bigg)\nn\\
&=\frac{2^{2+2\epsilon}}{\pi} \Gamma(\epsilon-3/2) (1+D_\alpha+D_\alpha^2) \Bigg[\frac{\Gamma(1+\alpha+\epsilon)}{\Gamma(5+\alpha)\Gamma(2\epsilon-2+\alpha)} (2z)^{1+\alpha+\epsilon}\bigg]_{\alpha=0}
\end{align}

The last line of~\eqref{eq:cnD6R41} involving the $\ell$-sum takes the form
\begin{align}
&\quad-2^{2+2\epsilon}\frac{\Gamma(\epsilon-3/2)}{\pi} \sum_{\ell\geq 0} \left( \frac{\Gamma(4+\ell-\epsilon) (2z)^{4+\ell-\epsilon}}{\Gamma(8+\ell-2\epsilon)\Gamma(1+\ell)}  - \frac{\Gamma(4+\ell+\epsilon) (2z)^{4+\ell+\epsilon}}{\Gamma(8+\ell)\Gamma(1+\ell+2\epsilon)} \right)\nn\\
&= -2^{2+2\epsilon}\frac{\Gamma(\epsilon-3/2)}{\pi}\frac{1}{1-D_\alpha} \bigg[ \frac{\Gamma(4+\alpha-\epsilon)(2z)^{4+\alpha-\epsilon} }{\Gamma(8+\alpha-2\epsilon)\Gamma(1+\alpha)}  - \frac{\Gamma(4+\alpha+\epsilon)(2z)^{4+\alpha+\epsilon} }{\Gamma(8+\alpha)\Gamma(1+\alpha+2\epsilon)} \bigg]_{\alpha=0}\,.
\end{align}
In this way, the whole seed Fourier mode $c_n(y)$ can be written as (finite) linear combinations of terms of the form $\sigma_a(n) (4\pi|n|)^b y^r e^{-2\pi|n| y}$ together with the action of differential operators on them, where we recall that $z=2\pi|n| y$.

The notation we introduce for this rewriting is
\begin{align}
\label{eq:cnD6R4split}
c_n(y) = c_n^{(1)}(y) + c_n^{(2)}(y)
\end{align}
where 
\begin{align}
c_n^{(1)} (y) &\label{eq:diffOP2}= \mathcal{D}^{(1)} \left(\sigma_{-2}(n) (4\pi |n|)^{-1-\epsilon}  (2z)^{1+\alpha+\epsilon}e^{-z} \right) \,,\\
c_n^{(2)} (y) &\nn= -\frac{2^{5+2\epsilon}\zeta(3+2\epsilon)\Gamma(\epsilon-3/2)}{\sqrt{\pi}(1-4\epsilon^2)}\nn\\
&\nn\hspace{5mm}\times \frac{1}{1-D_\alpha} \bigg[ \sigma_{-2}(n) (4\pi |n|)^{-1-\epsilon} e^{-z}\left(\frac{\Gamma(4+\alpha-\epsilon)(2z)^{4+\alpha-\epsilon} }{\Gamma(8+\alpha-2\epsilon)\Gamma(1+\alpha)}  - \frac{\Gamma(4+\alpha+\epsilon)(2z)^{4+\alpha+\epsilon} }{\Gamma(8+\alpha)\Gamma(1+\alpha+2\epsilon)}\right) \bigg]_{\alpha=0}
\end{align}
and
\begin{align}
\label{eq:diffOP1}
\mathcal{D}^{(1)} f(\alpha) &= -\frac{8\zeta(3+2\epsilon)}{\sqrt{\pi}(1-4\epsilon^2)}\frac{\partial_\alpha }{1-D_\alpha}\left[\left(\alpha^2-13\alpha+2 -2 \epsilon (\alpha^2+3\alpha+10) \right) \frac{\Gamma(\alpha+\tfrac12)}{\Gamma(5+\alpha)\Gamma(\alpha-1)} f(\alpha)\right]_{\alpha=0}\nn\\
&\hspace{10mm} +\frac{2^{5+2\epsilon}\zeta(3+2\epsilon)\Gamma(\epsilon-3/2)}{\sqrt{\pi}(1-4\epsilon^2)} (1+D_\alpha+D_\alpha^2) \Bigg[\frac{\Gamma(1+\alpha+\epsilon)}{\Gamma(5+\alpha)\Gamma(2\epsilon-2+\alpha)} f(\alpha)\bigg]_{\alpha=0}\,.
\end{align}
The reason that we have split the Fourier mode $c_n(y)$ in this way is because the two parts have different contributions to the asymptotic expansion.

\subsection{Applying the general formula for the asymptotic expansion}
\label{sec:AppAsymp}
We now apply the general formula~\eqref{eq:IgenAsy} for the asymptotic expansion of Fourier modes of the type~\eqref{eq:FMgen} to~\eqref{eq:cnD6R4split} to obtain the asymptotic expansion for $z\to 0$ in a similar decomposition
\begin{align}
\label{eq:D6R4asy}
I \sim I^{(1)} + I^{(2)} \,,
\end{align}
where, after exchanging the differential operator and the asymptotic expansion,
\begin{align}
\label{eq:DI1}
I^{(1)} = \mathcal{D}^{(1)} I(-2,\alpha,1+\alpha+\epsilon)
\end{align}
will be evaluated momentarily and $I^{(2)}$ follows from applying~\eqref{eq:IgenAsy} to $c_n^{(2)}(y)$. 

We begin by showing that $I^{(2)}$ vanishes in the asymptotic expansion for $\epsilon\to 0$. Writing out the geometric series $1/(1-D_\alpha) = \sum_{\ell \geq 0} D_\alpha^\ell$ again, a single term in $c_n^{(2)}(y)$ contributes to the asymptotic expansion either as $I(-2,3+\ell-2\epsilon,4+\ell-\epsilon)$ or $I(-2,3+\ell,4+\ell+\epsilon)$ which are given by
\begin{align}
&\nn I(-2, 3+\ell-\epsilon\pm\epsilon, 4+\ell\pm\epsilon)= \frac{2^{1-\epsilon}\pi y^{-\epsilon}}{\Gamma(4+\ell\pm \epsilon)}\Bigg[ \frac{y}{\pi} \frac{\Gamma(4+\ell-\epsilon\pm \epsilon)\Gamma(3+\ell+\epsilon \pm \epsilon)}{\Gamma(\epsilon)} 
 + \\
 &\nn+\frac{\pi}{y} \frac{\Gamma(2+\ell-\epsilon\pm \epsilon)\Gamma(5+\ell+\epsilon\pm \epsilon)}{\Gamma(2+\epsilon)}\frac{\zeta(2+2\epsilon)\zeta(3)}{\zeta(3+2\epsilon)} + \left(\frac{\pi}{y}\right)^{3+\ell-\epsilon\pm \epsilon} \sum_{n\geq 0} \left(\frac{-\pi}{y}\right)^n \frac{\Gamma(7+2\ell+n\pm 2\epsilon)}{n!\cdot \Gamma(4+\ell+n\pm \epsilon)} \\
 &\times \frac{\zeta(-3-\ell-n +\epsilon\mp\epsilon)\zeta(-1-\ell-n+\epsilon\mp \epsilon) \zeta(4+\ell+n+\epsilon \pm \epsilon)\zeta(6+\ell+n+\epsilon\pm \epsilon)}{\zeta(8+2\ell+2n\pm 2\epsilon)\zeta(3+2\epsilon)} \Bigg]\,.
\end{align}
Inspecting this expression we see that all individual terms are continuous and finite in the limit $\epsilon\to0$ (since $\ell\geq 0$ and $n\geq 0$), and we can set $\epsilon=0$; the very first term even vanishes. Since the two terms with $\pm \epsilon$ appear with opposite signs in $c_n^{(2)}(y)$ their contribution to the (perturbative) asymptotic behaviour vanishes when $\epsilon\to 0$ and thus:
\begin{align}
I^{(2)} \underset{\epsilon\to 0}{\sim} 0
\end{align}
and we are left with the contribution $I^{(1)}$ in~\eqref{eq:DI1} coming from $c_n^{(1)}(y)$.

In order to evaluate~\eqref{eq:DI1} we first note that
\begin{align}
\label{eq:ID6R4}
&\hspace{10mm}I(-2,\alpha,1+\alpha+\epsilon) \\
&= \frac{2^{1-2\epsilon} \pi y^{-\epsilon}}{\Gamma(1+\alpha+\epsilon)} \Bigg[ \frac{y}{\pi} \frac{\Gamma(1+\alpha)\Gamma(\alpha+2\epsilon)}{\Gamma(\epsilon)}\frac{\zeta(2+2\epsilon)\zeta(3)}{\zeta(3+2\epsilon)}
 -\frac1{12} \frac{\pi}{y} \frac{\Gamma(\alpha-1)\Gamma(2+\alpha+2\epsilon)}{\Gamma(2+\epsilon)} \frac{\zeta(2+2\epsilon)}{\zeta(3+2\epsilon)}\nn\\
&\hspace{3mm}+\left(\frac{\pi}{y}\right)^{\alpha} \sum_{n\geq 0} \left(\frac{-\pi}{y}\right)^n \frac{\Gamma(1+2\alpha+2\epsilon+n)}{n!\cdot\Gamma(1+\alpha+\epsilon+n)}\nn\\
&\hspace{40mm}\times  \frac{\zeta(2-\alpha-n)\zeta(-\alpha-n)\zeta(3+\alpha+2\epsilon+n)\zeta(1+\alpha+2\epsilon+n)}{\zeta(2+2\alpha+2\epsilon+2n)\zeta(3+2\epsilon)}\Bigg]\,.\nn
\end{align}

We begin with the contributions from the second line of the differential operator in~\eqref{eq:diffOP1}. These can be evaluated fully since 
\begin{align}
D_\alpha^\ell \left[\frac{\Gamma(1+\alpha+\epsilon)I(-2,\alpha,1+\alpha+\epsilon)}{\Gamma(5+\alpha)\Gamma(2\epsilon-2+\alpha)}\right]_{\alpha=0} = \frac{\Gamma(1+\ell+\epsilon)I(-2,\ell,1+\ell+\epsilon)}{\Gamma(5+\ell)\Gamma(2\epsilon-2+\ell)}\,,
\end{align}
and the factor $1/\Gamma(2\epsilon-2+\ell)$ goes to zero when $\epsilon\to 0$ and $\ell=0,1,2$. Therefore, we only need to analyse the potentially diverging terms for $\epsilon\to 0$ in $I(-2,\ell,1+\ell+\epsilon)$. Inspecting~\eqref{eq:ID6R4}, there are no such terms for $\ell=1$ and $\ell=2$. For $\ell=0$, there is such a term when $n=0$ in the sum and this is the only possible contribution. The result is that the second line of~\eqref{eq:diffOP1} contributes as 
\begin{align}
\label{eq:2ndline}
\lim_{\epsilon \to 0} \bigg( 2^6 \zeta(3) \Gamma(-3/2)\sqrt{\pi} \frac{\zeta(0)\zeta(1+2\epsilon)}{\Gamma(5)\Gamma(2\epsilon-2)}\bigg) = -\frac{32}9 \pi \zeta(3)
\end{align}
to the asymptotic expansion of $I^{(2)}$, i.e. to the $y$-independent terms. 

We are then left with the first line of the differential operator~\eqref{eq:diffOP1} acting on $I(-2,\alpha,1+\alpha+\epsilon)$ given in~\eqref{eq:ID6R4}. We treat the three terms in $I(-2,\alpha,1+\alpha+\epsilon)$ separately as they contribute at specific orders in $y$: The first term only contributes at order $y^1$, the second term only at order $y^{-1}$ and the third term in~\eqref{eq:ID6R4} contributes at all orders $y^{-k}$ for $k\geq 0$ (and potentially $\log(y)$). 

In order to evaluate the action of the differential operator on the first term in~\eqref{eq:ID6R4}, we observe that for cancelling the $1/\Gamma(\epsilon)$ in this expression, one requires $\alpha=0$ in the numerator of~\eqref{eq:diffOP1} and thus should not apply any of shift operators $D_\alpha^\ell$ contained in the differential operator, otherwise the result vanishes for $\epsilon\to 0$. The complete contribution at linear order in $y$ in the limit is therefore
\begin{align}
\label{eq:firstterm}
y^1 : \quad - 16  \zeta(2)\zeta(3)  \frac{y}{\sqrt{\pi}}  \lim_{\epsilon \to 0}  \partial_\alpha\bigg[ (\alpha^2-13\alpha+2) \frac{\Gamma(\alpha+1/2)\Gamma(1+\alpha) \Gamma(\alpha+2\epsilon)}{\Gamma(\alpha-1)\Gamma(5+\alpha)\Gamma(\epsilon)}  \bigg]_{\alpha=0} = \frac{2}{3} \zeta(2)\zeta(3) y\,.
\end{align}
This deals completely with the first term in~\eqref{eq:ID6R4}.

Proceeding to the second term in~\eqref{eq:ID6R4} we are dealing with the order $y^{-1}$. Since the potentially diverging $\Gamma(\alpha-1)$ cancels between~\eqref{eq:ID6R4} and the differential operator, we are left with 
\begin{align}
\label{eq:2ndterm}
&\hspace{2mm} \frac{2^{2-\epsilon}\pi^{3/2} y^{-1-\epsilon} }{3(1-4\epsilon^2)} \frac{\partial_\alpha }{1-D_\alpha} \bigg[\left(\alpha^2-13\alpha+2-2\epsilon(\alpha^2+3\alpha+10)\right)\frac{\Gamma(\alpha+\tfrac12)\Gamma(2+\alpha+\epsilon)\zeta(2+2\epsilon)}{\Gamma(1+\alpha+\epsilon)\Gamma(5+\alpha)\Gamma(2+\epsilon)}\bigg]_{\alpha=0}\nn\\
&\underset{\epsilon\to 0}{\longrightarrow}\frac{4\pi^{3/2}\zeta(2)}{3} y^{-1} \partial_\alpha \frac{1}{1-D_\alpha} \bigg[\left(\alpha^2-13\alpha+2\right)\frac{\Gamma(\alpha+1/2)\Gamma(2+\alpha)}{\Gamma(1+\alpha)\Gamma(5+\alpha)}\bigg]_{\alpha=0}\nn\\
&= \frac{4\pi^{3/2}\zeta(2)}{3} y^{-1} \partial_\alpha \bigg[ \frac{2\alpha (\alpha-2)(\alpha+1) \Gamma(\alpha+1/2)}{\Gamma(4+\alpha)}\bigg]_{\alpha=0}\nn\\
&= -\frac{40\zeta(4)}{3}   y^{-1}
\end{align}
The limit $\epsilon\to 0$ can be taken first since there are no singularities in this limit and the geometric sum converges uniformly. This is the complete contribution from the second term in~\eqref{eq:ID6R4}.

The final contribution we need to evaluate is the third sum term in~\eqref{eq:ID6R4} when the first line of the differential operator~\eqref{eq:diffOP1} is applied to it. This can have contributions at orders $y^{-k}$ for $k\geq 0$ and the lowest few values of $k$ have to be analysed differently from the generic cases due to diverging terms. 

Let us denote by
\begin{align}
\label{eq:un}
u_n = (-1)^n \left(\frac{\pi}{y}\right)^{\alpha+n} \frac{\Gamma(1+2\alpha+n + 2\epsilon)\zeta(2-\alpha-n)\zeta(-\alpha-n) \zeta(3+\alpha+n+2\epsilon)\zeta(1+\alpha+n+\epsilon)}{n! \cdot\Gamma(1+\alpha+n+\epsilon)\zeta(2+2\alpha+2n +2\epsilon)\zeta(3+2\epsilon)}
\end{align}
the $n$th term in the sum of the third term in~\eqref{eq:ID6R4}. We note that most terms depend only on $\alpha+n$ that also sets the order in $y$. Potential divergences in $u_n$ come from the zeta functions when the argument approaches one. This can happen for $\zeta(2-\alpha-n)$ when $(\alpha,n)=(1,0)$ or $(\alpha,n)=(0,1)$ and for $\zeta(1+\alpha+n+\epsilon)$ when $(\alpha,n)=(0,0)$. Shifts in $\alpha$ come from expanding out the geometric sum $1/(1-D_\alpha)=\sum_{\ell\geq 0} D_\alpha^\ell$. In the differential operator~\eqref{eq:diffOP1} there is also a factor $1/\Gamma(\alpha-1)$ that can vanish for $\alpha=0$ or $\alpha=1$ and combine with diverging terms in $u_n$.

At order $y^0$ and for the first line of the differential operator there is only a contribution from the $u_0$ in~\eqref{eq:ID6R4} and one should not apply any shift operator $D_\alpha$ for this order in $y$. This contribution has a divergent contribution in the limit $\epsilon \to 0$ and comes out as
\begin{align}
\label{eq:thirdterm0}
y^{0} : \quad \frac{32}9\pi \zeta(3) - \frac{\pi^{5/2} \Gamma(\epsilon+1/2) \zeta(1+2\epsilon)\zeta(3+2\epsilon)}{(9-6\epsilon)\Gamma(1+\epsilon)\zeta(2+2\epsilon)} y^{-\epsilon} +O(\epsilon)
\end{align}
that has to be combined with~\eqref{eq:2ndline} and thus cancels the finite piece. We also see that there is an explicit divergent piece due to the $\zeta(1+2\epsilon)$ that we leave as it is since it combines with a piece coming from $c_0(y)$. 

At order $y^{-1}$, there are two contributions coming from the third term in~\eqref{eq:ID6R4}, one with $u_1$ and no $\alpha$-shift and one with $u_0$ and a single $\alpha$ shift. Combining the two leads to\footnote{Here, as elsewhere, potential $\log(y)$ terms cancel.}
\begin{align}
\label{eq:thirdterm1}
y^{-1} : \quad \frac{52\zeta(4)}{3} y^{-1}
\end{align}
that has to be combined with the contribution~\eqref{eq:2ndterm}.

At order $y^{-2}$, there are three contributions from the third term in~\eqref{eq:ID6R4}. Combining these leads to 
\begin{align}
\label{eq:thirdterm2}
y^{-2} : \quad-\frac{\pi \zeta(3)^2 \zeta(5)}{4\zeta(6)} y^{-2}\,.
\end{align}

At order $y^{-3}$, there are four contributions from the third term in~\eqref{eq:ID6R4}. Combining these leads to 
\begin{align}
\label{eq:thirdterm3}
y^{-3} : \quad\frac{4\zeta(6)}{27} y^{-3}\,.
\end{align}

In order to analyse the general term $y^{-k}$ with $k\geq 4$ we note first that there are no potential singularities when $\epsilon\to 0$, so we take this limit first. Then we are left with evaluating for $k\geq 4$
\begin{align}
y^{-k} : \quad -16 \sqrt{\pi}\zeta(3) \partial_\alpha \bigg[ \sum_{\ell=0}^k D_\alpha^\ell \left((\alpha^2-13\alpha+2) \frac{\Gamma(\alpha+1/2)}{\Gamma(5+\alpha)\Gamma(\alpha-1)\Gamma(\alpha+1)} u_{k-\ell} \right)\bigg]_{\alpha=0}\,.
\end{align}
Inspecting~\eqref{eq:un} we see that the only the combination $\alpha+n$ appears in the zeta functions. Therefore all terms in the inner sum have the same common factor
\begin{align}
\frac{\zeta(2-\alpha-k)\zeta(-\alpha-k) \zeta(3+\alpha+k)\zeta(1+\alpha+k)}{\zeta(2+2\alpha+2k)}\,.
\end{align}
If $k$ is even, this function starts at order $\alpha^2$ when expanded around $\alpha=0$ due to the vanishing of the zeta function at negative even integers. Since the sum multiplying this common factor is regular at $\alpha=0$ this shows that there are no contributions at order $y^{-k}$ for $k=2n\geq 4$ with $n$ an integer. For odd $k$ the quotient of the zeta functions starts at order $\alpha^0$.

If $k$ is odd, we then look in more detail at the non-zeta factors that multiply the common zetas
\begin{align}
\left(\frac{\pi}{y}\right)^{\alpha+k} \sum_{\ell=0}^k (-1)^{k-\ell}  \frac{((\alpha+\ell)^2-13(\alpha+\ell)+2)\Gamma(\alpha+\ell+1/2)}{\Gamma(5+\alpha+\ell)\Gamma(\alpha+\ell-1)\Gamma(\alpha+\ell+1)} 
\frac{\Gamma(1+2\alpha+k+\ell)}{\Gamma(k-\ell+1) \Gamma(1+\alpha+k)} \,.
\end{align}
We have verified that this sum starts at order $\alpha^2$ for odd $5\leq k\leq 99$ and are confident that this holds for all odd $k\geq 5$. This means that there are no terms in the asymptotic expansion of the form $y^{-k}$ with $k>3$.

Let us collect all terms in the asymptotic expansion of $I$ as defined in~\eqref{eq:D6R4asy}. Combining the terms~\eqref{eq:2ndline},~\eqref{eq:firstterm},~\eqref{eq:2ndterm},~\eqref{eq:thirdterm0},~\eqref{eq:thirdterm1},~\eqref{eq:thirdterm2} and~\eqref{eq:thirdterm3} leads to
\begin{align}
\label{eq:Itotal}
I &\sim \frac23\zeta(2)\zeta(3) y + 4\zeta(4) y^{-1} -\frac{\pi \zeta(3)^2\zeta(5)}{4\zeta(6)} y^{-2} + \frac{4\zeta(6)}{27} y^{-3}\nn\\
&\hspace{10mm} -\frac{\pi^{5/2} \Gamma(\epsilon+1/2) \zeta(1+2\epsilon)\zeta(3+2\epsilon)}{(9-6\epsilon)\Gamma(1+\epsilon)\zeta(2+2\epsilon)} y^{-\epsilon} + O(\epsilon)\,.
\end{align}


\end{document}